\documentclass{svjour3}                     
\smartqed  
\usepackage{graphicx}
\usepackage{bm} 
\newcommand{\enbox}[1]{\enskip\hbox{#1}\enskip}
\newcommand{\DART}{DART\textsuperscript{\rm\textregistered}}

\usepackage{url}
\journalname{}
\begin{document}

\title{Detiding \DART\ buoy data for real-time extraction of source coefficients for operational tsunami forecasting
}

\titlerunning{Detiding \DART\ buoy data for operational tsunami forecasting}        

\author{Donald~B.~Percival \and
        Donald~W.~Denbo \and
        Marie~C.~Ebl{\'e} \and
        Edison~Gica \and
        Paul~Y.~Huang \and
        Harold~O.~Mofjeld \and
        Michael~C.~Spillane \and
        Vasily~V.~Titov \and
        Elena~I.~Tolkova
}


\institute{Donald B.~Percival \at
            Applied Physics Laboratory,
            University of Washington,
            Seattle, WA  98195-5640
            USA \\
            Department of Statistics,
            University of Washington,
            Seattle, WA  98195-4322
            USA \\
              \email{dbp@apl.washington.edu}           
           \and
           Donald~W.~Denbo \and Marie~C.~Ebl{\'e} \and Edison~Gica \and Harold~O.~Mofjeld \and Michael~C.~Spillane \and Vasily~V.~Titov \at %
           NOAA/Pacific Marine Environmental Laboratory,
           7600 Sand Point Way NE,
           Seattle, WA  98115
           USA 
           \and
           Donald~W.~Denbo \and Edison~Gica \and Harold~O.~Mofjeld \and Michael~C.~Spillane \and Vasily~V.~Titov \at
           Joint Institute for the Study of the Atmosphere and Ocean, University of Washington, Seattle, WA 98195-5672 USA
           \and
           Paul~Y.~Huang \at
           National Tsunami Warning Center,
           National Weather Service,
           Palmer, AK 99645
           USA   
           \and
           Elena~I.~Tolkova \at
           NorthWest Research Associates
           4126 148th Ave NE,
           Redmond, WA 98052
           USA
}

\date{Received: date / Accepted: date}

\maketitle

\begin{abstract}
U.S.~Tsunami Warning Centers use real-time bottom pressure (BP) data
transmitted from a network of buoys deployed in the Pacific and Atlantic Oceans
to tune source coefficients of tsunami forecast models.
For accurate coefficients and therefore forecasts,
tides at the buoys must be accounted for.
In this study, five methods for coefficient estimation are compared,
each of which accounts for tides differently.
The first three subtract off a tidal prediction based on
(1)~a localized harmonic analysis involving 29 days of data immediately
preceding the tsunami event,
(2)~68 pre-existing harmonic constituents specific to each buoy,
and (3)~an empirical orthogonal function fit to the previous 25 hrs of data.
Method~(4) is a Kalman smoother that uses method~(1) as its input.
These four methods estimate source coefficients after detiding.
Method~(5) estimates the coefficients simultaneously 
with a two-component harmonic model that accounts for the tides.
The five methods are evaluated using archived data
from eleven \DART\ buoys, 
to which selected artificial tsunami signals are superimposed.
These buoys represent a full range of observed tidal conditions
and background BP noise in the Pacific and Atlantic,
and
the artificial signals have a variety of patterns
and induce varying signal-to-noise ratios.
The root-mean-square errors (RMSEs) of least squares estimates of sources coefficients
using varying amounts of data are used to compare the five detiding methods.
The RMSE varies over two orders of magnitude between detiding methods,
generally decreasing in the order listed,
with method~(5) yielding the most accurate estimate of source coefficient.
The RMSE is substantially reduced
by waiting for the first full wave of the tsunami signal to arrive.
As a case study,
the five method are compared
using data recorded from the devastating 2011 Japan tsunami.
\end{abstract}

\section{Introduction}
\label{intro}
To collect data needed to provide coastal communities with timely tsunami warnings,
the National Oceanic and Atmospheric Administration (NOAA)
has deployed an array of Deep-ocean Assessment and Reporting of Tsunamis (\DART) buoys
at strategic locations in the Pacific and Atlantic Oceans
(Gonz{\'a}lez {\it et al.}, 2005;
Titov {\it et al.}, 2005;
Spillane {\it et al.}, 2008;
Mofjeld, 2009).
When a tsunami event occurs,
data from these buoys are analyzed at U.S.~Tsunami Warning Centers (TWCs)
using the Short-term Inundation Forecast for Tsunamis (SIFT) application
(Gica {\it et al.}, 2008; Titov, 2009).
The SIFT application was developed by the NOAA Center for Tsunami Research 
to rapidly and efficiently forecast tsunami heights at specific coastal communities.
SIFT compares \DART\ buoy data with precomputed models
as one step in creating the forecast.
Matching precomputed models with data
requires that
any tidal components in the data be
either removed or compensated for in some manner,
an operation that we refer to as detiding.
To facilitate the operational needs of SIFT,
detiding of data from \DART\ buoys must be done
as soon as possible after the data become available.

In the course of developing the SIFT application,
we have entertained multiple methods
for detiding \DART\ buoy bottom pressure (BP) data nearly in real time.
In this paper
we compare five such methods using archived data collected
by eleven \DART\ buoys.
These buoys are deployed in both the Pacific and Atlantic Oceans
in places with different tidal regimes.
Besides the dominant tides,
the archived data contain other BP fluctuations of the kind
that would be present during an actual tsunami event.
We take data from periods when no known significant tsunamis occurred
and introduce an artificial tsunami signal.
Each signal is patterned after a precomputed computer model
for an actual tsunami event.
The magnitude of the artificial event is controlled
by a source coefficient $\alpha$.
We consider five different methods
for extracting the artificial tsunami signal,
each of which handles the tidal component in a different manner.
We assess how well each of the five detiding methods allows
us to extract the known $\alpha$.
By repeating this scheme for many different combinations
of \DART\ data and artificial tsunami signals,
we can evaluate how well the five detiding methods work under idealized conditions
(thus, while we make use of observed tidally-dominated data,
we do not take into consideration confounding factors,
an important one being a mismatch between the actual tsunami event and our model for it).

The remainder of the article is organized as follows.
In Sect.~\ref{sec:BPmeasurements}
we review the format of BP measurements from \DART\ buoys
as received by U.S.~TWCs.
We next describe construction of simulated tsunami events,
which are formed by adding archived data from eleven representative buoys
to associated models for tsunami signals
(Sect.~\ref{sec:Scenarios}).
We give details about the five methods for extracting tsunami signals
in Sect.~\ref{sec:DetidingPressureMeasurements}.
The first two are well-known methods based on harmonic models.
The next two are non-standard linear filters
that utilize either empirical orthogonal functions
or Kalman smoothing in conjunction with a local-level state space model.
The final method, which is based on a regression model
that includes terms for both the tsunami signal and tidal components,
proves to be the method of choice
in study described in Sect.~\ref{sec:results}.
We compare the five methods using \DART\ buoy data
collected during the devastating 2011 Japan tsunami
in Sect.~\ref{sec:Tohoku}.
We state our conclusions and discuss our results
in Sect.~\ref{sec:conclusions}.


\section{Bottom pressure measurements from \DART\ buoys}
\label{sec:BPmeasurements}

A \DART\ buoy actually consists of two units:
a surface buoy and a unit located at the bottom of the ocean with a pressure recorder
(Gonz{\'a}lez {\it et al.}, 2005; NOAA Data Management Committee, 2008; Mofjeld, 2009).
The bottom unit stores measurements of water pressure
integrated over non-overlapping 15-sec time windows,
for a total of $60 \times 4 = 240$ measurements every hour.
These internally recorded measurements only become fully available
when the bottom unit returns to the surface and is recovered for servicing
(time between servicings can be as long as two or three years).
We refer to the internally recorded data as the 15-sec stream.
Normally the buoy operates in standard reporting mode
in which the bottom unit packages
together one measurement every 15 minutes
(a 60 fold reduction in data)
over a 6-h block
for transmission via acoustic telemetry
up to the surface buoy once every 6~h.
The surface buoy then relays the data up to the Iridium Satellite System
for dissemination to the outside world.
We refer to data collected in standard reporting mode as the 15-min stream.

When a typical tsunami event occurs,
the bottom unit detects a seismic event,
which causes the \DART\ buoy to go into event reporting mode.
The transmission of the 15-min stream is suspended
while the \DART\ buoy is in event reporting mode.
As part of this mode,
the bottom unit averages together four consecutive 15-sec measurements
and transmits these averages up to the surface buoy.
We refer to these data as the 1-min stream.
An additional 2 hours of 1-min data are transmitted on the hour
during the event reporting mode.
When the outside world first gets access to the 1-min stream,
there can be a gap between it and the most recent value of the 15-min stream
ranging up to almost 6~h.
After the on-hour data transmission
an additional 1 to 2 hours of 1-min data before the event will be available.
If we assume that tsunami events commence at random within the 6-h reporting
cycle for the 15-min stream,
then the average size of this gap will be 3~h.
The data that are thus typically available
during a tsunami event are a portion of the 15-min stream prior to the event
and a 1-min stream that becomes available piece-by-piece in real time
as the event evolves
and that contains at least 1~h of data occurring before the event.

\section{Scenarios and artificial tsunami signals}
\label{sec:Scenarios}

The goal of this paper is to objectively compare different detiding methods
for extracting tsunami signals in near real-time.
A model for what is recorded by a \DART\ buoy during a tsunami event has three components:
a tsunami signal, tidal fluctuations and background noise.
The latter is due to seismic, meteorological, measurement and other nontidal effects
(Cartwright {\it et al.}, 1987;
Niiler {\it et al.}, 1993;
Webb, 1998;
Cummins {\it et al.}, 2001;
Mofjeld {\it et al.}, 2001;
Zhao and Alford, 2009).
The purpose of detiding is to compensate for tidal fluctuations
in the recorded data (Consoli {\it et al.}, 2014).
Although \DART\ buoys have recorded a number of tsunami events,
use of these data to evaluate different detiding methods is problematic:
we ideally need to know the true tidal fluctuations,
and these are not known to sufficient accuracy for actual data
due to the presence of background noise.
One solution is to simulate each of the three components.
In combining these components to simulate tsunami events, 
we would then know the true tidal fluctuations
and thus be in a position to quantify
how well different detiding methods did;
however, simulating tidal fluctuations and background noise in a manner
that does not give an unfair advantage to certain detiding methods is tricky.
We can bypass simulation of both tidal fluctuations and background noise
by making use of actual BP measurements from \DART\ buoys
under the realistic assumption
that significant tsunami signals are rare events. 

\begin{figure}[t!]
  \noindent\includegraphics[width=28.5pc,angle=0]{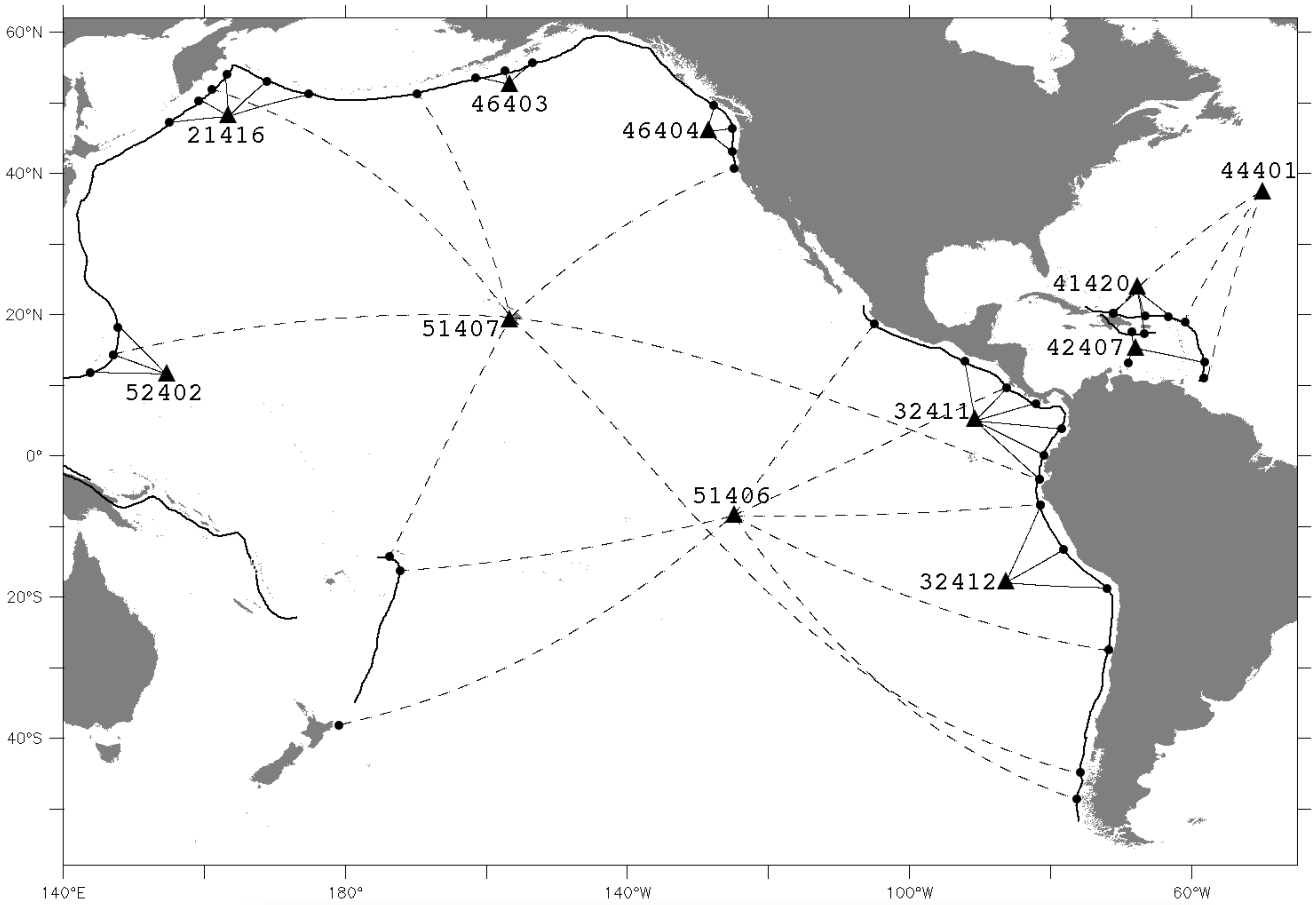}\\
  \caption{Locations of eleven \DART\ buoys (triangles) and 42 unit sources (solid circles).  All but two of the unit sources are located in subduction zones,
which are shown by solid curves
(the remaining two are near New Zealand and Venezuela in areas
where the plate boundary does not have subduction characteristics,
but is nonetheless capable of lesser seismic events).
The unit sources that are used with \DART\ buoys 44401, 51406, and 51407 are remote,
and dashed curves are used to show the buoy/unit source pairings via a Great Circle;
the pairings associated with the remaining eight buoys are linked by solid lines.
  }
  \label{fig:DetidingPaperChart}
\end{figure}

In this paper we use archived 15-sec streams retrieved
from eleven representative \DART\ buoys.
The buoys are in the Atlantic and Pacific Oceans
and are listed in Table~\ref{tab:buoys}.
Their locations are displayed in Fig.~\ref{fig:DetidingPaperChart}.
These particular buoys were chosen
because the data they collect represent four well-known types of tides,
hence offering the study of detiding over a broad range of ocean tidal conditions.
All data were obtained from NOAA's National Geophysical Data Center (NGDC;
the reader is referred to Mungov, {\it et al.}, 2012, 
for details about data collection and processing).
The units for the data archived by NGDC are in pounds per square inch absolute (psia),
which we convert to water depth in meters by multiplication by $0.67$~m/psia
(this conversion factor is based on a standard ocean,
but its value does not impact any of the results we present).

\begin{figure}[t!]
  \noindent\includegraphics{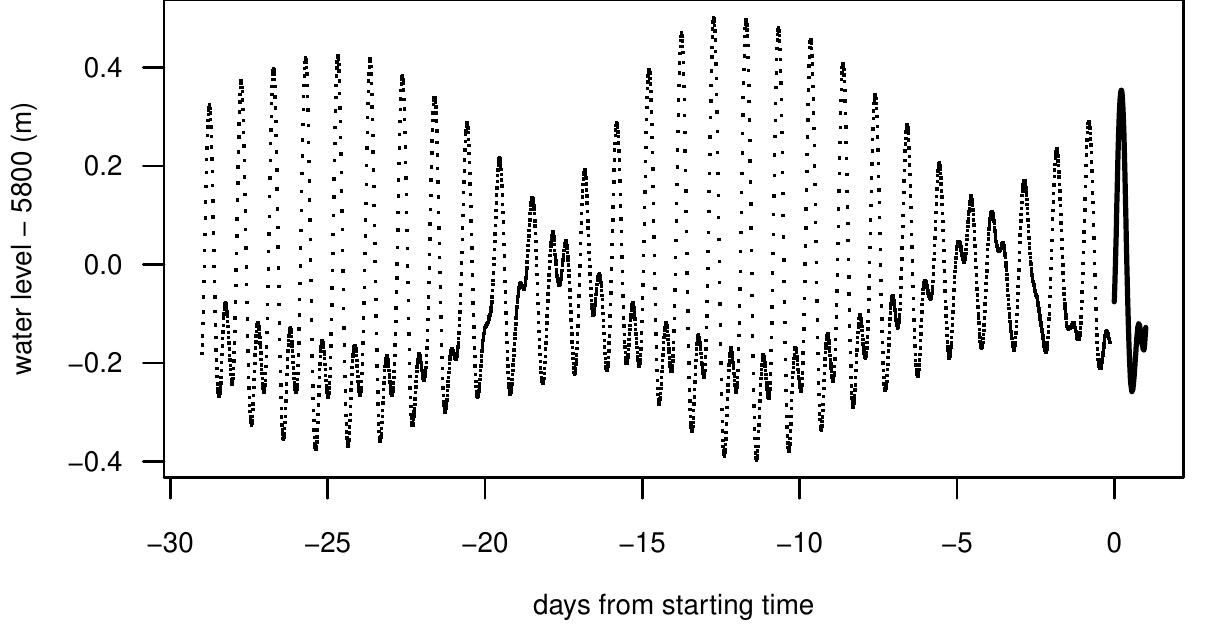}\\
  \caption{Scenario 943 constructed from data recorded by \DART\ buoy 52402
with starting time $t_0$ set to 9:21:00~UT on 27 June 2007. 
The points show the 15-min stream, which starts 29 days prior to $t_0$,
but has a 3-h gap just prior to it.
The dark curve on the right-hand side is the 1-min stream, which starts at $t_0$
and ends one day later.
  }
  \label{fig:scenario}
\end{figure}

As documented in Table~\ref{tab:buoys},
we have from 321 to 998 days of 15-sec streams for each of the eleven buoys.
We use these streams to construct `scenarios'
mimicking the tidal fluctuations and background noise that might have been present
in 15-min and 1-min streams available during an actual tsunami event.
To do so for a particular buoy,
we start by selecting a random starting time $t_0$.
The first part of the scenario associated with $t_0$
consists of a 29-day segment of a 15-min stream
extracted by subsampling from the 15-sec stream immediately prior to $t_0$.
To mimic typical operational conditions,
we create a 3-h gap prior to $t_0$
by eliminating part of the constructed 15-sec stream.
Using data from the 15-sec stream occurring immediately after $t_0$,
we form a 1-day segment of a 1-min stream by averaging four adjacent values.
The constructed 15-min and 1-min streams constitute one scenario.
For each buoy we form a total of 1000 scenarios
with starting times $t_0$ chosen at random without replacement
from the set of all possible starting times
(for example, as indicated in Table~\ref{tab:buoys} for buoy 21416,
this set ranges from 29 days after 07/25/2007 up to 1 day prior to 07/01/2009).
Figure~\ref{fig:scenario} shows an example of one scenario.
(There are occasional missing values in the 15-sec streams archived by NGDC.
These can lead to gaps in a constructed 15-min stream
in additional to the usual 3-h gap prior to $t_0$.
All of the detiding methods in our study can handle these gaps.
If, however, the missing values lead to gaps in the constructed 1-min stream,
we have elected to disregard the chosen $t_0$ and randomly pick a new one
merely to simplify evaluation of the performance of the various detiding methods.
Thus, a given scenario can have gaps at arbitrary times in its 15-min stream,
but none in its 1-min stream.)

With tidal fluctuations and background noise
being handled using actual data from \DART\ buoys,
we now turn our attention to simulating tsunami signals
and combining these with the other two components
to simulate the 1-min streams observed during a tsunami event.
Let $\bar{\bf y}$ be a vector
containing a simulated 1-min stream
observed during a tsunami event starting at time $t_0$
and lasting for one day
(the bar over ${\bf y}$ is a reminder
that the 1-min stream is produced from four-point averages
of the 15-s stream).
We model this vector as 
\begin{equation}\label{eq:modelK}
\bar{\bf y}
=
{\bf x} + \alpha_1 {\bf g}_1 + \cdots +  \alpha_K {\bf g}_K
+ {\bf e},
\end{equation}
where ${\bf x}$, ${\bf g}_1$, \dots, ${\bf g}_K$ and ${\bf e}$
are vectors representing,
respectively, tidal fluctuations, components derived from $K\ge1$ unit sources
(these collectively serve to model the tsunami signal) 
and background noise,
while $\alpha_1$, \dots, $\alpha_K$ are $K$ nonnegative scalars
known as source coefficients (Percival {\it et al.}, 2011). 
Forming the ${\bf g}_k$'s is discussed in detail in
Titov {\it et al.}~(1999) 
and Gica {\it et al.}~(2008), 
and the following overview is taken from
Percival {\it et al.}~(2011). 
Tsunami source regions are defined along subduction zones and other portions of oceans
from which earthquake-generated tsunamis are likely to occur
(see the solid curves on Fig.~\ref{fig:DetidingPaperChart}).
Each source region is divided up into a number of unit sources,
each of which has a fault area of $100\times 50$~$\hbox{km}^2$.
For each pairing of a given \DART\ buoy with a particular unit source,
a model ${\bf g}_k$ has been calculated predicting
what would be observed at the buoy from time $t_0$ and onwards
based upon the assumption that
the tsunami event was caused by 
a reverse-thrust earthquake
with standardized moment magnitude $M_W=7.5$
starting at $t_0$
and located within the unit source.
The source coefficient $\alpha_k$ is used
to adjust the standardized magnitude in the model
to reflect the magnitude in an actual tsunami event.
The sum of the $\alpha_k$'s is a measure of the overall strength
of the tsunami event and provides initial conditions
for models that predict coastal inundation.
As discussed in Percival {\it et al.}~(2011), 
given $K$ candidate unit sources
and based on varying amounts of data,
a least squares procedure can be used to obtain
statistically tractable estimates of the $\alpha_k$'s
under the assumption that
the tidal fluctuations ${\bf x}$
are known to reasonable accuracy.

\begin{figure}[t]
  \noindent\hskip-6pt\includegraphics{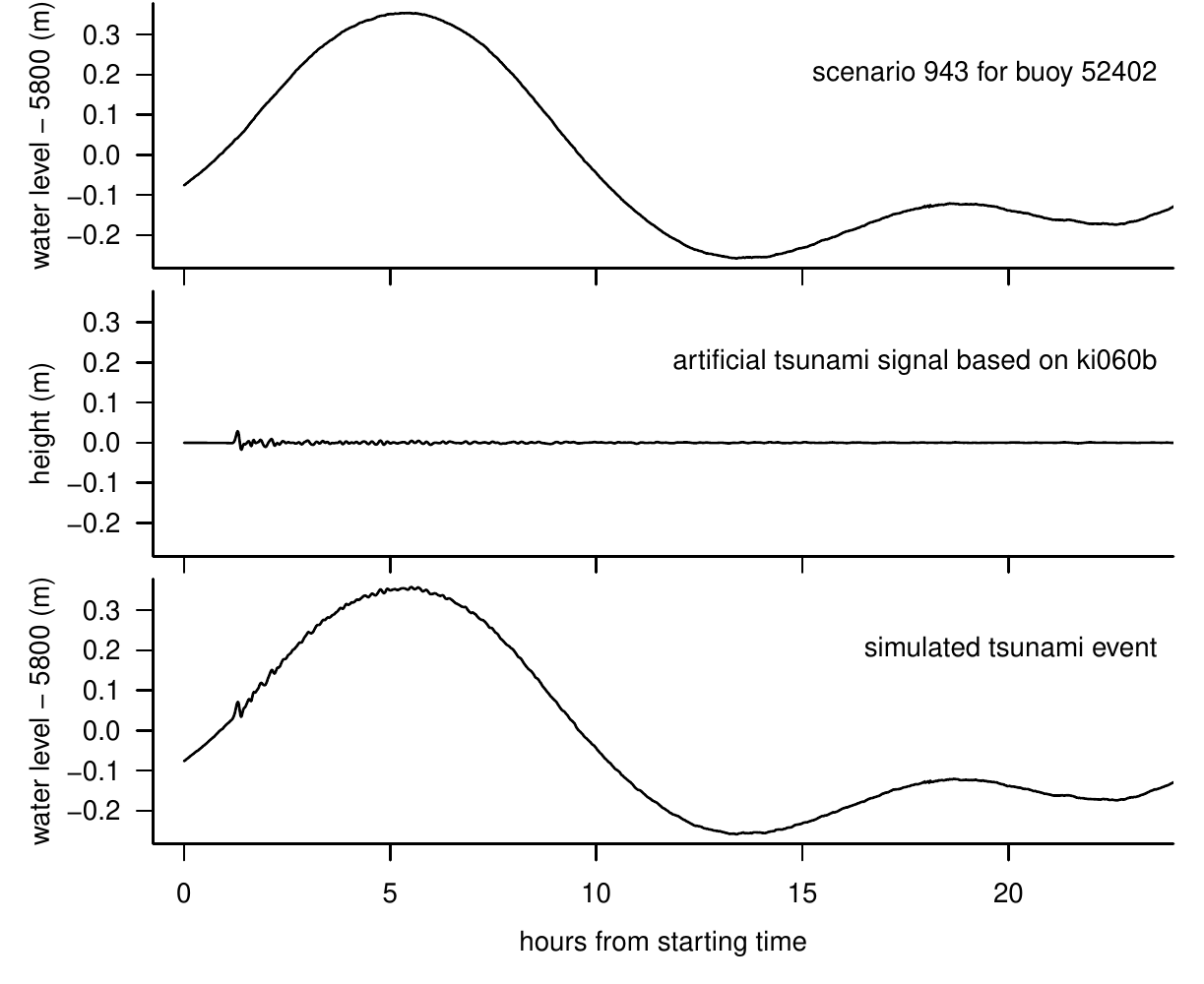}\\
  \caption{Construction of simulated tsunami event.
The upper plot is the 1-min stream from a scenario
for \DART\ buoy 52402 (this is shown in Fig.~\ref{fig:scenario}
as the dark curve on the right-hand side).
This stream is assumed to consist of tidal fluctuations
and background noise summed together, i.e., ${\bf x} + {\bf e}$
from Equation~\ref{eq:model1}.
The middle plot shows $\alpha{\bf g}$ with $\alpha$ set to 6,
where $\bf g$ is based on ki060b,
one of the unit sources associated with 52402 (see Table~\ref{tab:unitSources}).
The bottom plot is a simulated tsunami event $\bar{\bf y}$ formed by
adding $\alpha {\bf g}$ to ${\bf x} + {\bf e}$.
  }
  \label{fig:oneEvent}
\end{figure}

To focus on detiding,
we assume a simplified version of Equation~\ref{eq:modelK},
namely,
\begin{equation}\label{eq:model1}
\bar{\bf y}
=
{\bf x} + \alpha {\bf g} + {\bf e};
\end{equation}
i.e., the simulated 1-min stream 
involves just a single unit source and a single source coefficient.
The scenarios for a given buoy are stand-ins for ${\bf x} + {\bf e}$.
We take the unit source-based ${\bf g}$
and multiply it by $\alpha$ to
form an artificial tsunami signal $\alpha {\bf g}$.
From an examination of actual tsunami events,
we set $\alpha = 6$ as a representative source coefficient.
Figure~\ref{fig:oneEvent} shows an example of constructing
a simulated tsunami event based upon the 1-min stream
shown in Fig.~\ref{fig:scenario}
and a unit source chosen for \DART\ buoy 52402.

\begin{figure}[t]
  \noindent\hskip-25pt\includegraphics{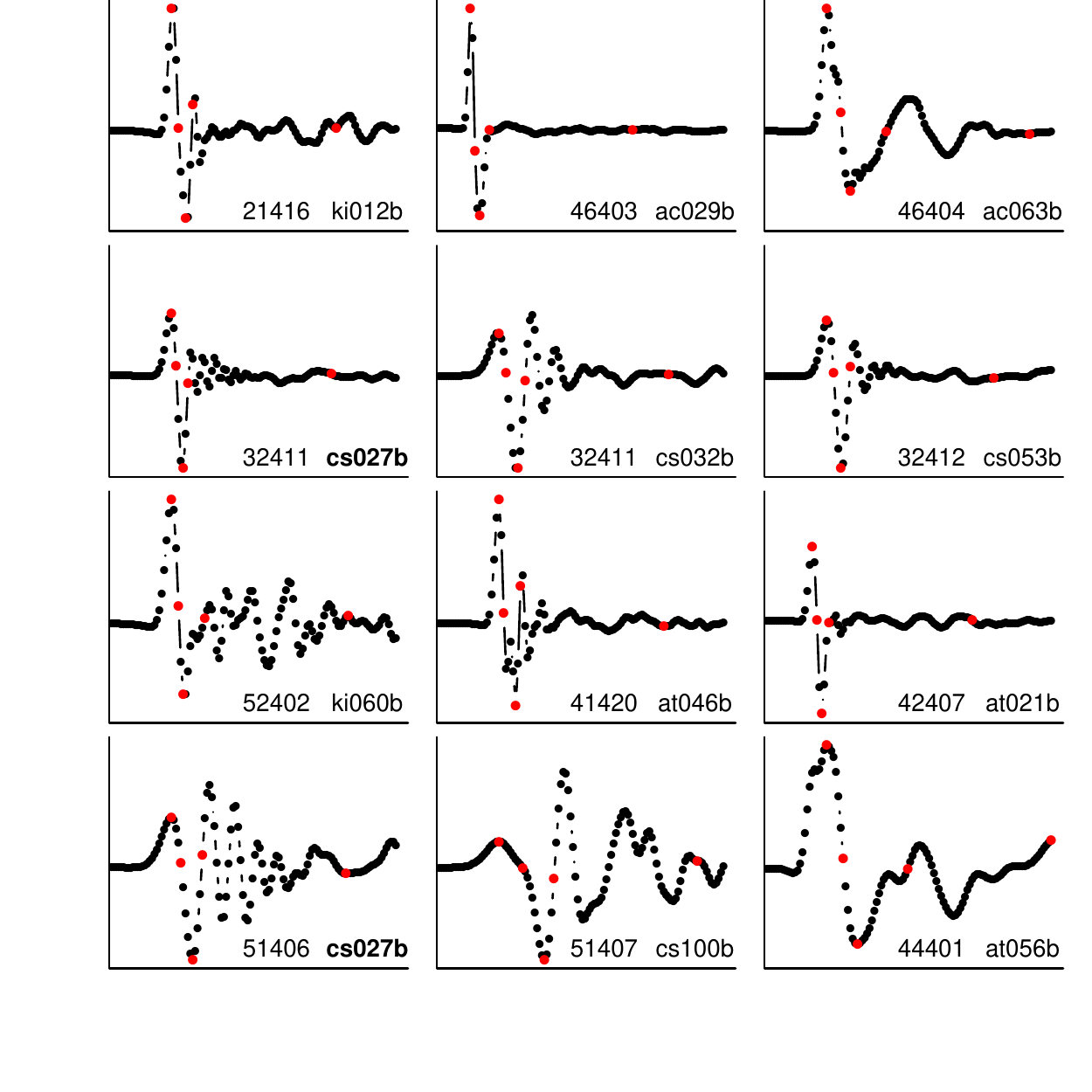}\\
  \caption{Twelve artificial tsunami signals $\alpha \bf g$.
Each signal is based upon a model for what would be observed eventually
over a 120-min segment of time
at a particular buoy due to an earthquake originating within a particular unit source
(the starting time of each segment is different for each signal).
The source coefficient $\alpha$ is adjusted separately for each signal
merely for plotting purposes
(the actual ranges for the unscaled $\bf g$'s are listed in Table~\ref{tab:unitSources}).
There are five hand-picked red points in each plot.
These are associated with varying amounts of data
related to the first full wave of the signal,
namely, the first quarter of the full wave, half, three-quarters,
all the full wave and one-hour past the end of the first full wave.
  }
  \label{fig:unitSources}
\end{figure}

For each of the eleven buoys in our study,
we picked from three to seven unit sources
with different orientations with respect to the buoy --
these choices are listed in Table~\ref{tab:unitSources}
and depicted as solid circles in Fig.~\ref{fig:DetidingPaperChart}.
For \DART\ buoys near a subduction zone,
we selected the closest unit source
so that the buoy would be in the main beam of the radiated tsunami energy.
We then added two additional unit sources,
bearing 30 to 45 degrees to either side of the first,
to represent waves traveling obliquely toward the buoy.
In the case of \DART\ buoy 21416,
which is representative of the northwest Pacific
and proximate to both the Aleutian and Kamchatka source regions,
we selected unit sources from both regions.
Buoy 32411 (located close to both Central and South America)
has a similar geometry,
so we chose three unit sources from Central America
and three from South America.
For mid-ocean buoys (51406, 51407 and 44401),
we selected unit sources from several subduction zones
to which the buoy might respond.
There are 42 unit sources in all, 5 of which are used by two buoys,
for a total of 47 pairings of buoys and unit sources.
Each pairing leads to a different artificial tsunami signal $\alpha\bf g$. 
Twelve representative signals are shown in Fig.~\ref{fig:unitSources}
(here the coefficient $\alpha$ is adjusted separately for each signal
merely for plotting purposes
-- 
the actual range for all 47 $\bf g$'s is listed in Table~\ref{tab:unitSources}
and varies from 0.2 to 13.1~cm). 
Each plot shows a 120-min segment of a given artificial tsunami signal $\alpha\bf g$,
one point for each minute,
but a different segment for each signal.
For use later on,
five hand-picked points are colored red.
From left to right, the first four of these points
mark the approximate occurrence of a quarter,
a half, three-quarters and all of the first full wave
comprising the signal
(Table~\ref{tab:unitSources} lists the actual times associated
with these points).
The final (right-most) point marks one hour past
the end of the first full wave.

Each of the eleven buoys is represented once in Fig.~\ref{fig:unitSources},
except for buoy 32411, which has two signals associated with it
that are visually quite different
(left-hand and middle plots on second row).
One of these two depicted signals
(cs027b, left-hand plot, second row)
is generated from the same $100\times 50$~$\hbox{km}^2$ region
as the signal depicted for 51406 (left-hand plot, bottom row).
Again these two signals are visually quite different.
These duplicate-buoy/duplicate-unit-source pairings illustrate the fact
that each signal depends upon both the location of the buoy
and the location of the unit source.

\begin{figure}[t]
  \noindent\hskip-25pt\includegraphics{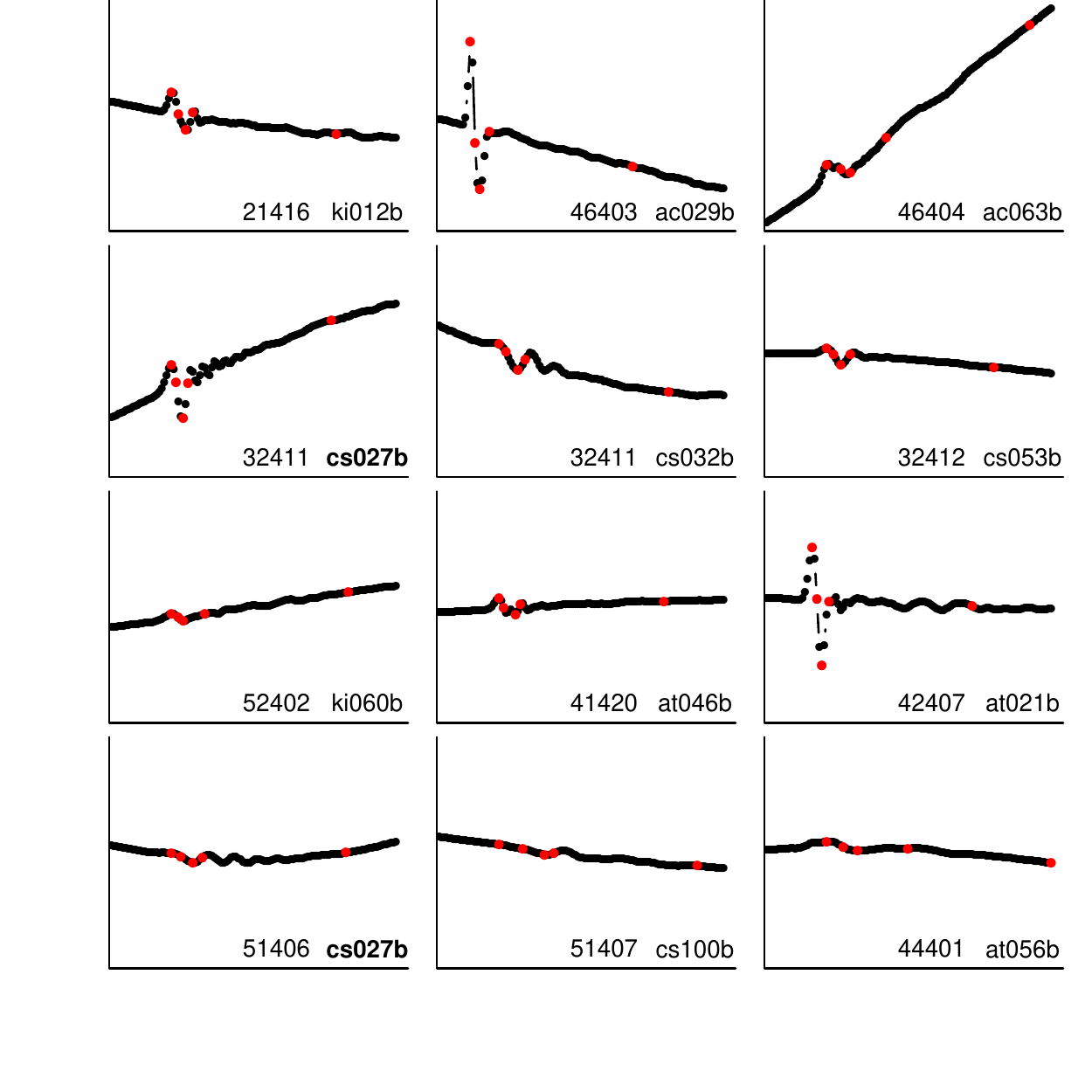}\\
  \caption{Twelve simulated tsunami events $\bf y$.
Each event is formed by adding
the artificial tsunami signal $\alpha \bf g$
shown in Fig.~\ref{fig:unitSources} (but with $\alpha$ now set to 6)
to the 1-min stream of a scenario randomly chosen from amongst
the 1000 scenarios for a given buoy.
The vertical height of each plot is 1.2m.
  }
  \label{fig:tsunamiEvents}
\end{figure}

Figure~\ref{fig:tsunamiEvents} shows 120-min segments of twelve simulated tsunami events,
each of which make use of the artificial tsunami
signals shown in corresponding plots of Figure~\ref{fig:unitSources}.
These events were formed by adding $\alpha \bf g$ (with $\alpha$ now set to 6)
to the 1-min stream from a randomly chosen scenario for each buoy
(the scenario chosen for buoy 52402 is the same one used in Fig.~\ref{fig:oneEvent}
-- thus the left-most plot in the third row of Fig.~\ref{fig:tsunamiEvents}
is a zoomed-in version of bottom plot of Fig.~\ref{fig:oneEvent}).
The vertical axis for each plot in Fig.~\ref{fig:tsunamiEvents} spans 1.2m.
It is easy to visually pick out the tsunami signal in some plots,
but harder in others,
which illustrates the fact that,
even though $\alpha=6$ in all cases,
the signal-to-noise ratio varies substantially.

\section{Methods for handling tides in bottom pressure measurements}
\label{sec:DetidingPressureMeasurements}
Here we describe five methods that take BP measurements $\bar{\bf y}$
and use them to estimate the source coefficient $\alpha$ in Equation~\ref{eq:model1}.
Each method deals with the tidal fluctuations $\bf x$ in a different manner.
The first four methods do so by predicting or estimating the fluctuations using,
say, $\hat {\bf x}$.
This prediction is then subtracted from $\bar{\bf y}$ to yield detided BP measurements:
\[
{\bf d} = \bar{\bf y} - \hat {\bf x} = \alpha {\bf g} + \bm{\epsilon},
\]
where $\bm{\epsilon} = {\bf e} + {\bf x}  - \hat {\bf x}$
is an error term encompassing both background noise and
inaccuracies in predicting the tidal fluctuations.
Given the detided measurements,
we then use the ordinary least squares (OLS) method to estimate $\alpha$, yielding the estimator
\begin{equation}\label{eq:alphaHat}
\hat \alpha = \frac{{\bf g}^T{\bf d}}{{\bf g}^T{\bf g}},
\end{equation}
where ${\bf g}^T$ denotes the transpose of the vector ${\bf g}$.
The fifth method is different from the other four
in that it uses OLS to create $\bf d$ and to estimate $\alpha$ jointly.
The inversion algorithms for estimating source coefficients
from multiple buoys
currently in SIFT
(Percival, {\it et al.}, 2011) 
and under development~(Percival, {\it et al.}, 2014) 
are based on least squares methods,
but go beyond OLS by including nonnegativity constraints
and penalties to induce automatic unit source selection.
These algorithms assume the model of Equation~\ref{eq:modelK}
for the multiple buoys,
but proceed under the assumption that the BP measurements
have been adjusted so that tidal fluctuations
have been removed as much as possible.
For studying how best to deal with tides,
it suffices to use a single buoy and the simplified model of Equation~\ref{eq:model1},
and there is no real gain in using anything
other than the OLS estimator of Equation~\ref{eq:alphaHat}.

The first two of the five methods are based on harmonic modeling,
which we describe in Sect.~\ref{sec:Harmonic}.
The third method is based on empirical orthogonal functions
(Sect.~\ref{sec:EOF}),
while the fourth employs Kalman smoothing,
but makes use of harmonic modeling for initialization purposes 
(Sect.~\ref{sec:KS}).
Sect.~\ref{sec:joint} describes the fifth method,
which in part involves a simplified version of harmonic modeling.

\subsection{Harmonic modeling methods}
\label{sec:Harmonic}

The harmonic prediction method is the standard one used by NOAA
to predict the tides at coastal stations. For detiding \DART\ data,
it can be thought of as the classic method of tidal prediction.
It assumes that the tides are sums of sinusoidal constituents,
each with its own frequency.
To make tidal predictions at a \DART\ station,
the amplitude and phase lag of each constituent 
are determined via a tidal analysis of observed BP data.

The first detiding method is a harmonic prediction method
that carries out the tidal analysis in the following manner.
Consider one of the scenarios described in Section~\ref{sec:Scenarios}.
Each scenario consists of a one-day segment of a 1-min stream starting at time $t_0$,
which is preceded by 29 days (less 3 hours) of a 15-min stream.
Without loss of generality,
set $t_0=0$ to simplify the discussion,
and let $y_n$ denote an observation from the 15-min stream,
but with $n$ indexing the underlying 15-sec stream
from which the 15-min stream is subsampled;
i.e.,
the actual time associated with $y_n$ is $n\,\Delta$, where $\Delta=15$s.
Assuming $y_n$ to be tidally dominated,
we entertain a harmonic model of the form
\begin{eqnarray}
y_n &=& \mu + \sum_{m=1}^M A_m \cos(\omega_m n\,\Delta - \phi_m) + e_n \label{eq:HarmModelAmpPhase} \\
&=& \mu + \sum_{m=1}^M \left[ B_m \cos(\omega_m n\,\Delta) + C_m \sin(\omega_m n\,\Delta)\right] + e_n \label{eq:HarmModelBC}
\end{eqnarray}
for $n = -780, -840, \ldots, -166920, -166980$
(note that index $n = -780$ corresponds to the last value in the observed 15-min stream,
which occurs prior to the 3-h gap,
while $n = -166980$ indexes the first value,
which occurs 29 days prior to $t_0=0$).
In the above $\mu$ is an unknown overall mean level;
$\omega_m$ is one of $M$ known tidal frequencies;
$B_m$ and $C_m$ are unknown coefficients
that can be used to deduce the amplitudes $A_m$ and phase lags $\phi_m$;
and $e_n$ is a residual term (hopefully small).
We use an OLS fitting procedure
to estimate $\mu$, $B_m$ and $C_m$
via, say, $\hat \mu$, $\hat B_m$ and $\hat C_m$.
If we replace $\mu$, $B_m$ and $C_m$ by their estimates,
we can then use the right-hand side of Equation~\ref{eq:HarmModelBC}
with the residual term set to zero to predict
what the tidal fluctuation should be at any desired time index $n$.
After an artificial tsunami signal is added
to the 1-min stream of the scenario to form a tsunami event,
we detide this stream by forming
\begin{equation}\label{eq:DetideOneMinuteUsingHA}
d_n =
\bar y_n - \hat \mu -
\frac{1}{4} \sum_{k=0}^3 \sum_{m=1}^M
\left[ \hat B_m \cos(\omega_m [n+k]\,\Delta) + C_m \sin(\omega_m [n+k]\,\Delta) \right]
\end{equation}
for $n=0, 4, \ldots, 5756, 5760$,
where $\bar y_n$ is an element from the 1-min stream $\bar{\bf y}$.
For 29 days of prior data,
we make use of $M=6$ tidal frequencies $\omega_1$, \dots, $\omega_6$
commonly referred to as N2, M2, S2, Q1, O1 and K1
(see, e.g., Table~1 in Ray and Luthcke, 2006). 
The $d_n$'s given above form the elements of the vector $\bf d$
used to form the estimator $\hat \alpha$ of Equation~\ref{eq:alphaHat}.

\begin{figure}[t]
  \noindent\hskip-6pt\includegraphics{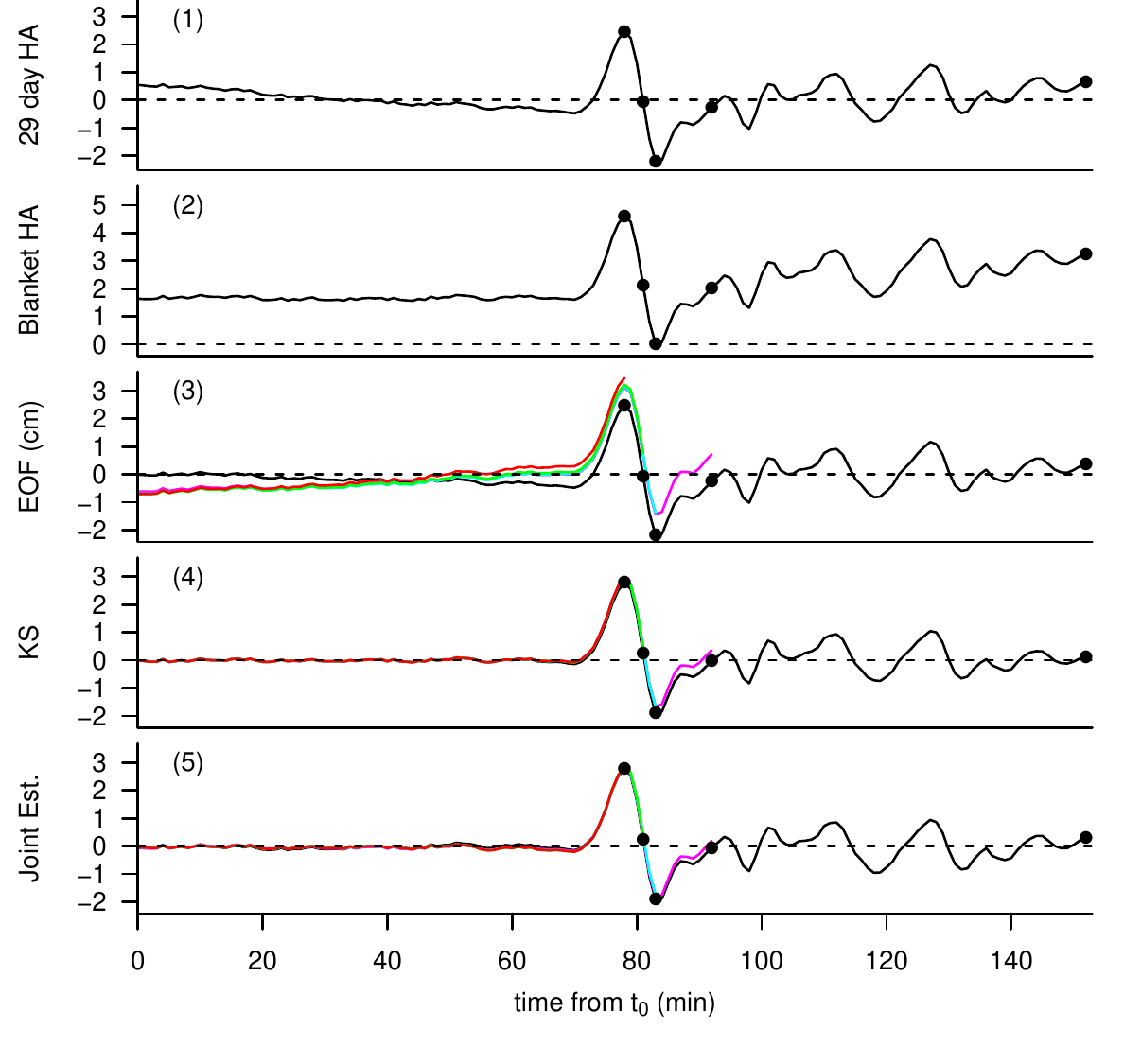}\\
\caption{Detided bottom pressure measurements for simulated tsunami event
shown in bottom plot of Fig.~\ref{fig:oneEvent}.
Each plot corresponds to one of the five methods under study.
From top to bottom,
these are
(1)~a 6 constituent harmonic analysis (HA)
based on 29 days of the 15-min stream before $t_0$ (the start of the 1-min stream),
(2)~a 68 constituent HA based on all available 15-sec data,
(3)~an empirical orthogonal function approach,
(4)~a Kalman smoother approach
and
(5)~a 2 constituent local HA with joint estimation of the source coefficient $\alpha$.
In plots~(3) to~(5),
there are five color-coded curves,
each corresponding to detiding based upon one of five different amounts of data
(some curves in~(4) and~(5) are partially obscured
because they are virtually identical to other curves).
Red indicates use of data from $t_0$ up to a quarter of the first full wave (FFW)
of the tsunami signal;
green, to half of the FFW;
cyan, to three quarters;
magenta, to the end of the FFW;
and black, to one hour past the end of the FFW.
The five ending times are shown by black circles
(the first four are listed in Table~\ref{tab:unitSources}
on the row pertaining to buoy 52402 and unit source ki060b,
and all five are shown as red circles 
in the left-hand plots on the third rows of Figs.~\ref{fig:unitSources}
and~\ref{fig:tsunamiEvents}). 
}
  \label{fig:detidingFive}
\end{figure}

The first row of Fig.~\ref{fig:detidingFive} shows
detided BP measurements $d_n$ corresponding 
to the simulated tsunami event
shown in the bottom plot of Fig.~\ref{fig:oneEvent}.
Visually there is evidence that this detiding is not entirely satisfactory:
there is a systematic drift downwards over the first hour
that arguably is due to tidal fluctuations
and thus should not be present in $d_n$.
As described in the caption to the figure,
the five black circles mark various time points
associated with the artificial tsunami signal.
If we place the data from $t_0$ up to one of these five time points
into the vector $\bf d$
and create the corresponding vector $\bf g$,
we can obtain an estimate of the source coefficient $\alpha$
using Equation~\ref{eq:alphaHat}.
The resulting estimates $\hat \alpha$
are listed in the first row of Table~\ref{tab:alphaEstimates}.
Recalling that the true value of $\alpha$ is 6,
we see that, not unexpectedly,
the better estimates are associated with larger amounts of data.
Looking at estimates based on varying amounts of data is
of considerable operational interest.
As more BP measurements become available as a tsunami event evolves,
we can expect in general to get better estimates of $\alpha$,
but at the expense of a delay in issuing timely warnings.
Determining how much an estimate of $\alpha$ is likely to improve
by waiting for more data
is vital for managing the trade-off between accuracy and timeliness.
For this example, there is improvement in waiting until the first full
wave occurs, but none in waiting an hour past that time.

The second detiding method is based on a harmonic analysis
that, for a given buoy, is based on {\it all\/}
the 15-s data listed for it in Table~\ref{tab:buoys}.
This `blanket' harmonic model has the same form as Equation~\ref{eq:HarmModelAmpPhase},
but now the time index $n$ increases in steps of one rather than 60,
and we use $M=68$ sinusoidal constituents. 
For optimal accuracy of this type of an analysis,
the measurements should span at least one year,
which Table~\ref{tab:buoys} indicates is true
for all buoys with the exception of 41420
(this buoy has 321 days of data, slightly less than a year).
The tidal predictions are made by adjusting the lunar harmonic constants
for perigean (8.85-year) and nodal (18.6-year) variations in the moon's orbit,
computing the height associated with each constituent at the times of interest,
and then summing these heights to yield the prediction.
Detiding of the 1-min series in a scenario is accomplished
using an equation similar to Equation~\ref{eq:DetideOneMinuteUsingHA},
the only difference being that the overall mean level $\mu$
is estimated using the scenario's 15-min stream
rather than being pre-specified.
A particular strength of the second detiding method is that
it requires minimal use of the 15-min stream
(it is only used to estimate $\mu$).
In one extreme case,
this stream was entirely missing for buoy 21416 over a six-week period
prior to 1/15/2009,
when a Kuril Islands event triggered reporting of the 1-min stream.
If the 15-min stream is not available,
it is possible to estimate $\mu$ using just the 1-min stream,
but care would be needed to ensure
that the tsunami signal does not unduly distort the estimate.

A number of software packages are available to do tidal analyses
of observations
and to make the tidal predictions,
a standard one being the Foreman FORTRAN 77 package
(Foreman, 1977, revised 2004). 
For our study, we used tidal predictions generated by NGDC  
(see Mungov {\it et al.}, 2012, 
for details).
 
The second row of Fig.~\ref{fig:detidingFive} shows
detided BP measurements $d_n$ produced by this second harmonic method.
In contrast to the first method,
we no longer see a systematic drift downwards over the first hour;
however, the $d_n$'s during that hour are systematically elevated above zero,
which is questionable.
The five estimates $\hat \alpha$ corresponding to different amounts of data
are listed in the second row of Table~\ref{tab:alphaEstimates}.
These estimates are worse than the ones we obtained
from the first harmonic method except when using the largest amount of data;
however, we shouldn't rely on this single example to draw conclusions
about the relative merits of the two harmonic methods
-- see the discussion in Section~\ref{sec:results}.

When tidal predictions are subtracted from BP measurements,
fluctuations always remain in the tidal bands.
They are due to non-stationary fluctuations, non-linear tides,
and tidal constituents not accounted for in the tidal analysis.
Of these, internal tides are certainly significant.
They are generated around the ocean margins and shallow ridges
and then propagate elsewhere in the ocean
(e.g., Cummins {\it et al.}, 2001; Zhou and Alford, 2009).  
The residual tides limit the degree to which the total tide can be removed
from BP data through simple subtraction of a predicted tide,
even when the tidal analysis is performed on the same time series
(as is the case here).

\subsection{Empirical orthogonal function method}
\label{sec:EOF}
The two best known approaches for detiding data
are to subtract off a prediction from a harmonic model
(as described in the previous section)
and to subject the data to a linear time-invariant (LTI) high-pass filter.
To isolate a tsunami signal without distortion,
the high-pass filter should retain components with periods as long as 2~hrs.
The so-called `edge effects' of such a filter distort
at least 1-hr sections at the beginning and at the end of the tsunami signal.
Thus an LTI filter cannot reliably isolate a tsunami signal
immediately after it is registered by a \DART\ buoy,
whereas it is desirable to make use of this data
as soon as possible following the start of a tsunami event.
Moreover, most digital filters are designed to work with regularly sampled data
and are not easy to adapt if there are gaps in the data.
While the BP unit in a \DART\ buoy internally records
a measurement once every 15~s,
only gappy segments of these data (or 1-min averages thereof)
are typically available externally following an earthquake event. 

In this section and the next, we explore two approaches
to detiding involving linear (but not time-invariant) filters.
These approaches are tolerant of data gaps and are less prone to edge effects,
thus overcoming the two problems we noted about standard LTI high-pass filters. 
The first approach is based on extracting the tidal component in a segment of data
by back and forth projection onto a specific sub-space
in an $N$-dimensional space of vectors.
The sub-space is spanned by the empirical orthogonal functions (EOFs)
of segments of length $N$ of archived 15-min streams
from \DART\ buoys (Tolkova, 2009; Tolkova, 2010).  
The following description of this approach is based on Tolkova~(2010),
to which we refer the reader for more details.

The EOF method relies on the premise
that, due to the structure of tides in the deep ocean,
the sub-space spanned by the leading EOFs of tidally dominated data segments
of up to 3-days in length is fairly universal across various \DART\ buoys.
Oceanic tidal energy is concentrated
in the long-period, diurnal, and semidiurnal frequency bands
centered around 0, 1, and 2 cycles per day (cpd).
The effective diurnal band is from 0.8 to 1.1~cpd,
and the semidiurnal, from 1.75 to 2.05~cpd,
so the bandwidth for both bands is 0.3~cpd
(Munk and Cartwright, 1966).  
Tidal motion thus has two inherent time scales:
one day (the apparent tidal quasi-period)
and 3.3~days
(the shortest tidal segment from which
we can in theory resolve individual constituents
within either of the two major bands).
Since tides at different locations differ
only in the fine structure of the tidal bands,
the premise behind the EOF method 
says that the sub-space spanned by the leading EOFs of tidal segments
is essentially the same at all \DART\ locations
as long as the segment length is so short
as to not allow resolution of the fine structure in the tidal bands.

\begin{figure}[t]
  \noindent\includegraphics{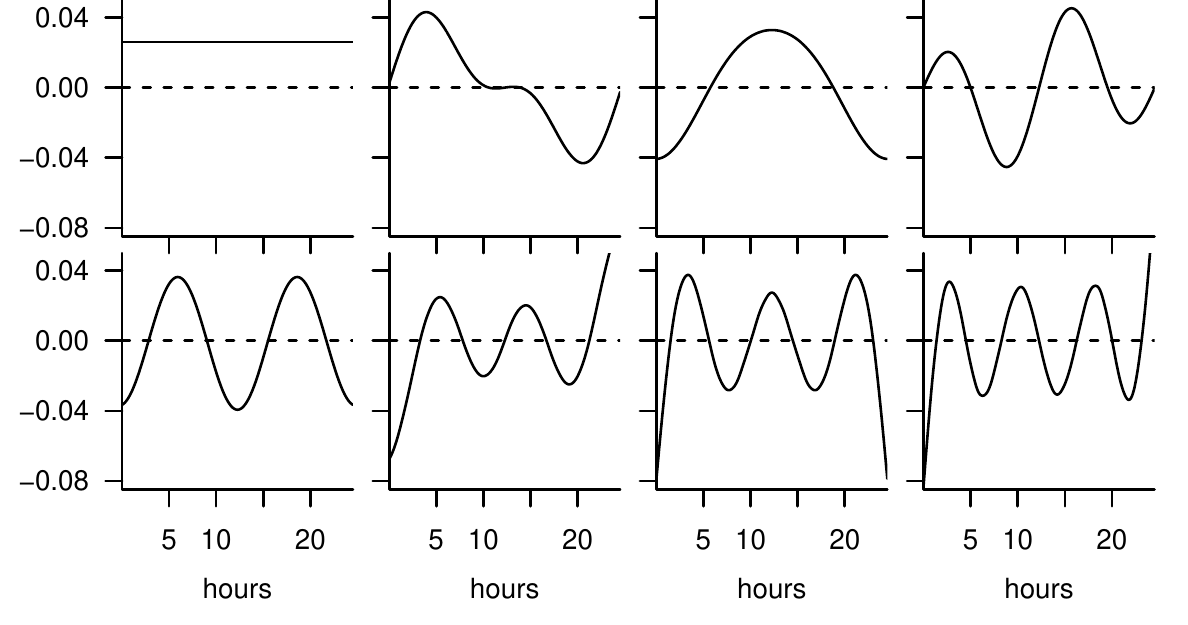}\\
  \caption{
Basis vectors ${\bf f}_m$, $m=0,1,\ldots,7$, used in EOF-based detiding.
The vectors consist of a constant vector ${\bf f}_0$
and vectors related to EOFs having the seven largest eigenvalues.
The vectors are arranged in the plot according to their number $m$ of zero crossings.
  }
  \label{fig:EightEOFs}
\end{figure}

Tolkova~(2010) computed the basis of the EOF sub-space of tidally dominated data
encompassing one lunar-day (24~h 50~min)
using an ensemble of 250 segments
taken from \DART\ buoy 46412 in 2007.  
Each segment consists of 99 readings sampled at 15~min intervals.
The EOFs associated with the seven largest eigenvalues,
complemented with a constant vector with elements all equal to $1/\surd99$,
provide the basis for a sub-space spanned by $M=8$ vectors of dimension $99$.
This sub-space is sufficient to capture the bulk of tidal variations
in segments of length 99 from a 15-min stream recorded by any \DART\ buoy.
To process 1-min streams within the SIFT system,
the $M$ vectors were re-sampled to a 1-min interval,
yielding vectors of length $N=15\times 98+1=1471$.
Re-orthogonalization and re-normalization
produces vectors ${\bf f}_m$, $m=0,1, \ldots, M-1$,
satisfying ${\bf f}^T_m{\bf f}_n=0$ for $m\not=n$
and $=1$ for $m=n$.
Figure~\ref{fig:EightEOFs} shows the resulting eight vectors. 

Three steps are needed to accomplish detiding using the ${\bf f}_m$ vectors.
For the sake of argument,
suppose the vector $\bar {\bf y}$ contains a segment
from a 1-min stream of length $N=1471$
(i.e., approximately one lunar day).
First, we project this segment onto the vectors ${\bf f}_m$
to obtain $M$ coefficients $c_m = {\bf f}_m^T \bar {\bf y}$.
Second, we estimate the tidal component by taking the $M$ vectors,
multiplying them by their corresponding coefficients
and then adding together the resulting scaled vectors.
Finally, this estimate of the tidal component is subtracted from $\bar {\bf y}$,
yielding the detided data $\bf d$.
Mathematically, we can write
\begin{equation}\label{eq:detideEOFregular}
{\bf d}=\bar {\bf y} - F{\bf c}
\enbox{with}
{\bf c}  = F^T \bar {\bf y},
\end{equation}
where $F$ is a $N \times M$ matrix
whose columns are ${\bf  f}_0, \dots, {\bf f}_{M-1}$,
and ${\bf c}$ is a vector containing the $M$ coefficients.
Letting $\bar y_n$ and $F_{m,n}$ denote, respectively,
the $n$th element of $\bar {\bf y}$
and the $(m,n)$th element of $F$
and momentarily regarding
\[
\sum_{n=0}^{N-1} \left( \bar y_n - \sum_{m=0}^{M-1} c_m f_{m,n} \right)^2
\]
as a function of $c_0, \ldots c_{M-1}$,
we note that setting $c_m$ equal to ${\bf f}_m^T \bar {\bf y}$
results in minimizing the above sum of squares. 

We can generalize this technique to handle the case of irregular sampling 
by the following simple procedure,
which ignores the distinction between the 1-min and 15-min streams.
For a span of $N=1471$ minutes of interest,
construct a vector $\tilde {\bf y}$ of length $N$
that contains all available values from either one of the streams.
Let $w_n=1$ if the $n$th element $\tilde y_n$ of $\tilde {\bf y}$
is actually available from one of the streams,
and let $w_n=0$ if it is not available.
We set $\bf c$ such that
\begin{equation}
\sum_{n=0}^{N-1} w_n \left( \tilde y_n - \sum_{m=0}^{M-1} c_m f_{m,n} \right)^2
\label{eq:aminim}
\end{equation}
is minimized with respect to $c_0, \ldots, c_{M-1}$.
Minimization of~(\ref{eq:aminim})
is a least squares problem
whose associated normal equations are 
\begin{equation}
H^T H {\bf c}=H^T \tilde {\bf y},
\label{eq:aequ}
\end{equation}
where $H$ is an $N \times M$ matrix
whose $(n,m)$th element is $w_n f_{m,n}$.
The system~(\ref{eq:aequ}) has a numerically viable solution
as long as the symmetric $M \times M$ matrix $H^TH$
is not poorly conditioned.
In this case the detided data are contained in
\begin{equation}
{\bf d}
= \tilde {\bf y} - H{\bf c}
\enbox{with}
{\bf c}
= (H^TH)^{-1}H^T\tilde {\bf y}.
\label{zfilter}
\end{equation}

The third row of Fig.~\ref{fig:detidingFive} shows
detided BP measurements $d_n$ produced by the EOF method.
In contrast to the two methods based on harmonic analysis,
the values that are subtracted from the 1-min stream
to accomplish EOF detiding depend upon the stream itself
and hence can change as different amounts of this stream
are utilized,
as this example illustrates.
In particular we see a systematic upward drift
when utilizing data less than or equal to the first full wave,
but this drift disappears
when we use data one hour past the end of the first full wave.
The five estimates $\hat \alpha$ corresponding to different amounts of data
are listed in the third row of Table~\ref{tab:alphaEstimates}.
For the three largest amounts of data,
these estimates are an improvement in this example over the ones
we obtained from the two harmonic methods.

\subsection{Kalman smoothing method}
\label{sec:KS}

Kalman smoothing (KS) has been used
in numerous applications as a method for
optimally smoothing time series
as new values of the series become available over time.
The optimality of this procedure is contingent
upon our ability to adequately describe
the underlying dynamics of the time series
in terms of a so-called state space model.
KS-based detiding has been advocated before
(see Consoli {\it et al.}, 2014, and references therein),
but the approach we describe here for detiding data from \DART\ buoys
differs from previous approaches in important aspects 
(for detailed expositions on KS, see
Brockwell and Davis, 2002;  
Durbin and Koopman, 2012;  
Shumway and Stoffer, 2011).  

Our KS approach is a two-stage procedure.
As before, let $\bar y_n$ represent the 1-min stream
from a given scenario
to which we have added an artificial tsunami signal.
The first stage is to detide $\bar y_n$ using the first harmonic modeling method,
yielding the detided series $d_n$, $n= 0, 4, 8, \ldots$,
via Equation~\ref{eq:DetideOneMinuteUsingHA}.
Merely to simplify equations that follow,
we reindex this first-stage detided series by defining 
$\tilde d_n = d_{4n}$, $n=0,1,2,\ldots$.
The second stage applies KS to $\tilde d_n$,
for which we assume a local level model (also called a random walk plus noise model;
see Brockwell and Davis, 2002;  
Durbin and Koopman, 2012).  
This model consists of two equations,
the first of which is known as the state equation,
and the second, the observation equation.
The state equation takes the form
\[
\mu_{n+1} = \mu_n + \zeta_n,\quad
n=0,1,2,\ldots,
\]
where $\mu_0$ is the initial state variable,
and $\zeta_n$ is a white noise sequence with mean zero and variance $\sigma^2_\zeta$.
The observation equation takes the form
\[
\tilde d_n  = \mu_n + \delta_n,\quad n=0,1,2,\ldots,
\]
where $\delta_n$ is another white noise sequence with mean zero,
but now with time-varying variance $\sigma^2_{\delta,n}$
(the sequences $\delta_n$ and $\zeta_n$ are uncorrelated).
The intent with this model is
to use $\mu_n$ to track any tidal component
left over in the first-stage detiding $\tilde d_n$
and to compensate for the presence of the tsunami signal
by adjusting $\sigma^2_{\delta,n}$ appropriately.
Based upon the initial state variable $\mu_0$,
the first-stage detided data $\tilde d_0, \ldots, \tilde d_n$
and the parameters $\sigma^2_\zeta$ and $\sigma^2_{\delta,n}$,
there is an elegant set of equations known as the Kalman recursions
that give the best (in the sense of minimum mean square error) linear
estimates of the unknown state variables $\mu_1, \ldots, \mu_n$. 
We denote these estimates as $\hat \mu_{1\vert n}, \ldots, \hat \mu_{n\vert n}$. 
For a fixed index $m\le n$,
the estimate $\hat \mu_{m\vert n}$ of $\mu_m$ changes as $n$ increases,
i.e., as more and more of the first-stage detided series becomes available.  
Given $\tilde d_0, \ldots, \tilde d_n$,
we define the KS-based detided series to be 
\[
\hat \delta_m = \tilde d_m - \hat \mu_{m\vert n},
\quad m = 0, 1, \ldots, n.
\]
The corresponding estimate of the source coefficient $\alpha$ is given by
Equation~\ref{eq:alphaHat},
where now the vector $\bf d$ contains the $\hat \delta_m$ variables.

The unknown parameters that we must set to implement KS-based detiding
are $\mu_0$, $\sigma^2_{\delta,n}$ and $\sigma^2_\zeta$.
For purposes of this paper,
we just use $\tilde d_0$ to estimate $\mu_0$;
however,
to offer some protection against rogue values,
the operational version of SIFT has an option for using the median of $\tilde d_0, \ldots, \tilde d_4$.
Our estimate $\hat \sigma^2_{\delta,n}$ of $\sigma^2_{\delta,n}$
is the sample variance of the seven variables $\tilde d_{n-3}, \ldots, \tilde d_{n+3}$,
with estimates for the first three variances
$\sigma^2_{\delta,0}$, $\sigma^2_{\delta,1}$ and $\sigma^2_{\delta,2}$
being set to $\hat \sigma^2_{\delta,3}$
(estimates for the last three variances are handled in an analogous manner).
We determined a setting for the final parameter $\sigma^2_\zeta$
in the following manner.
For all 1000 scenarios for a given pairing of a buoy and an artificial tsunami signal
with the source coefficient $\alpha=6$,
we used the KS detiding method
to compute source coefficient estimates $\hat \alpha_i$, $i=1,\ldots, 1000$,
over a grid of values for $\sigma^2_\zeta$ and for various amounts of data.
We then determined which value minimized $\sum_i ( \hat \alpha_i - 6)^2$.
Different pairings of buoys and signals and different data lengths
led to different minimizing values.
We set $\sigma^2_\zeta=6.25\times 10^{-13}$
after considering a large collection of representative pairings and data lengths --
while this value was not optimal for all such pairings and lengths,
it performed well overall.

The fourth row of Fig.~\ref{fig:detidingFive} shows
detided BP measurements $\hat \delta_n$ produced by the KS method.
The starting point for this method is the detided series
shown in the first row,
which was produced by the first harmonic modeling method.
A comparison of these two detided series
shows that the KS method has eliminated the downward trend
that is evident in the first seventy minutes
of the detrended series produced by harmonic modeling.
In contrast to the first three methods,
there is no evidence in the KS detided series
of a drift or offset that might be attributable
to lingering tidal fluctuations.
Similar to EOF detiding,
KS detided values can change as different amounts of the 1-min stream are utilized,
but the changes in the KS method are smaller
than those for the EOF method in this example.
The five KS-based estimates of $\hat \alpha$
corresponding to different amounts of data
are listed in the fourth row of Table~\ref{tab:alphaEstimates}.
For this example,
these estimates are closer to the true value $\alpha=6$
than the ones corresponding to the first three methods.


\subsection{Harmonic modeling method with joint source coefficient estimation}
\label{sec:joint}

The four detiding methods we have considered so far are similar in
that they all produce a detided series $\bf d$.
We then use $\bf d$ to produce an estimate $\hat \alpha$ of the source coefficient
via the OLS estimator of Equation~\ref{eq:alphaHat}.
The fifth and final method estimates the tidal component
jointly with the source coefficient based on just the 1-min stream
(merely to simplify the description below,
we assume this stream to be gap-free,
but it is easy to reformulate this method to handle gaps).
This joint estimation method is based on the model
\begin{equation}\label{eq:joint}
\bar {\bf y} = {\mu}{\bf 1}
+ \sum_{m=1}^2 \left(B_m {\bf {c}}_m + C_m {\bf {s}}_m\right)
+ \alpha {\bf g} + {\bf e}, 
\end{equation}
where $\bar {\bf y}$ is an $N$ dimensional vector containing
a portion of the 1-min stream from a scenario
to which the artificial tsunami signal $6{\bf g}$ has been added;
${\bf 1}$ is a vector of ones;
${\bf {c}}_m$ is a vector whose elements are $\cos\,(\omega_m n\,\Delta)$,
$n=0,\ldots,N-1$, with $\omega_2$ and $\omega_1$
being, respectively, the tidal frequency M2 and half that frequency
and with $\Delta = 1$~min;
${\bf {s}}_m$ is analogous to ${\bf {c}}_m$,
but with sines replacing cosines;
$\bf e$ is a vector of errors (presumed to have mean zero and a common variance);
and $\mu$, $B_1$, $C_1$, $B_2$, $C_2$ and $\alpha$
are unknown parameters.
In essence, this method estimates the tidal component
via a two-constituent harmonic model
(as alternatives to this model, we also considered harmonic models
with other than two constituents and polynomial models of various orders,
but the two-constituent harmonic model worked best overall
for joint estimation of $\alpha$).
Equation~\ref{eq:joint} can be rewritten as
\begin{equation}\label{eq:jointRewritten}
\bar {\bf y} = X{\bm \beta} + {\bf e}, 
\end{equation}
where $X$ is an $N\times 6$ design matrix whose columns
are ${\bf 1}$, ${\bf {c}}_1$, ${\bf {s}}_1$, ${\bf {c}}_2$, ${\bf {s}}_2$
and ${\bf g}$,
while ${\bm \beta} = (\mu, B_1, C_1, B_1, C_1, \alpha)^T$ is a vector of coefficients.
The OLS estimate
$\hat {\bm \beta} = (\hat\mu, \hat B_1, \hat C_1, \hat B_1, \hat C_1, \hat\alpha)^T$
of ${\bm \beta}$
satisfies the normal equations
\begin{equation}\label{eq:normalEquations}
X^T X \hat{\bm\beta} = X^T \bar{\bf y}
\enbox{and hence}
\hat{\bm\beta} = (X^T X)^{-1} X^T \bar{\bf y},
\end{equation}
subject to the invertibility of $X^T X$
(in the study discussed in Sect.~\ref{sec:results},
no instances of non-invertibility were encountered).
We can take the detided series for this method to be
\begin{equation}\label{eq:jointDetided}
{\bf d} = \bar {\bf y} - \hat{\mu}{\bf 1}
- \sum_{m=1}^2 \left(\hat B_m {\bf {c}}_m + \hat C_m {\bf {s}}_m\right).
\end{equation}
A strength of this detiding method is that,
in contrast to the other four methods,
it does not make any use of the 15-min stream,
which, as we've noted before, was entirely missing
in one actual tsunami event.  

The fifth row of Fig.~\ref{fig:detidingFive} shows
detided BP measurements ${\bf d}$ produced by the joint estimation method,
with the five estimates of $\hat \alpha$ corresponding to different amounts of data
being listed in the fifth row of Table~\ref{tab:alphaEstimates}.
Although in this example the detided series for this and the KS method
are visually similar to each other,
the joint estimation-based $\hat \alpha$ estimates
are always closer to the true value of $\alpha=6$
than the KS-based estimates.

%
%

\section{Comparison of five detiding methods}\label{sec:results}

Here we compare the five detiding methods described in the previous section
by considering how well each method estimates the source coefficient $\alpha$
from simulated tsunami events constructed as per Equation~\ref{eq:model1}
(an example of one such event is shown in Fig.~\ref{fig:oneEvent}).
For each of the 47 buoy/unit source pairings listed in Table~\ref{tab:unitSources},
we constructed 1000 simulated events
based upon the 1000 scenarios created for each buoy.
For each such event and for five different amounts of data from the 1-min stream
ranging from a quarter of the first full wave (1/4 FFW) of the tsunami signal
up to an hour past the end of the FFW,
we estimated $\alpha$ using the five detiding methods,
thus yielding $47\times1000\times5=235,000$
estimated coefficients for each method.
The task at hand is to summarize how well each method did.

\begin{figure}[t]
  \noindent\hskip-6pt\includegraphics{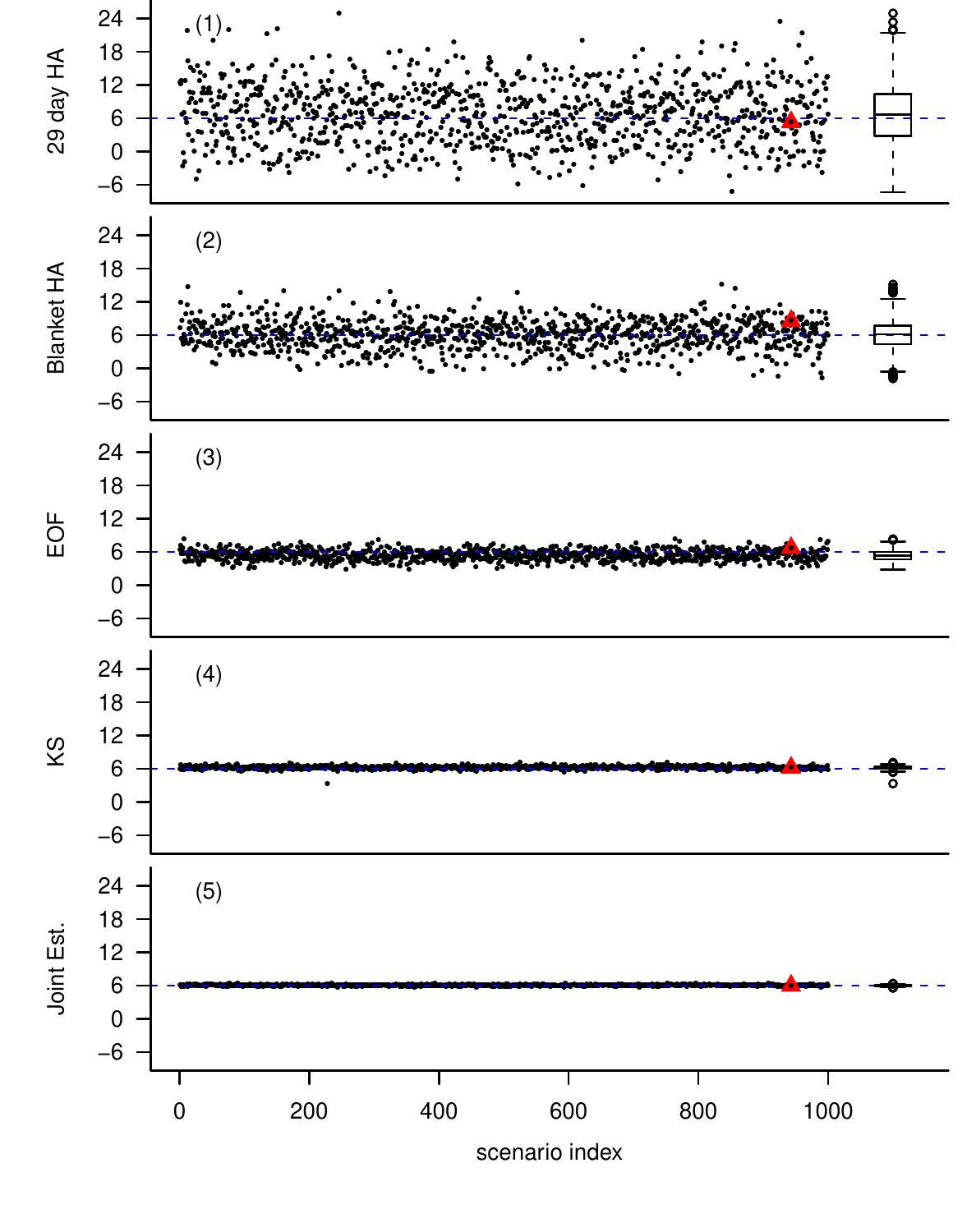}\\
\caption{
Estimated source coefficients $\hat \alpha$ (dots) given by,
from top to bottom,
(1)~a 6 constituent harmonic analysis (HA)
based on 29 days of the 15-min stream before $t_0$ (the start of the event),
(2)~a 68 constituent HA based on all available 15-sec data,
(3)~an empirical orthogonal function approach,
(4)~a Kalman smoother approach
and
(5)~a 2 constituent local HA with joint estimation of the source coefficient $\alpha$.
Each estimate is based on data up to three quarters of the first full wave
from one of the 1000 scenarios for buoy 52402,
to which has been added an artificial tsunami signal
based on unit source ki060b with source coefficient $\alpha=6$.
The dots for the estimates for scenario 943 are surrounded by a red triangle
(this scenario is used as an example in Figs.~\ref{fig:scenario},
\ref{fig:oneEvent} and~\ref{fig:detidingFive}
and in Table~\ref{tab:alphaEstimates}).
The boxplot on each row summarizes the distribution of the corresponding $\hat \alpha$ estimates (see text for details).
}
  \label{fig:scatterFive}
\end{figure}

We start by considering results for the pairing of buoy 52402 with unit source ki060b 
(this same combination is used in all or part
of Figs.~\ref{fig:scenario} to \ref{fig:detidingFive}
and in Table~\ref{tab:alphaEstimates}).
Figure~\ref{fig:scatterFive} has five rows, one for each of the five methods.
The dots in a given row show the 1000 estimates $\hat \alpha$
derived from data just up to 3/4 FFW.
Ideally we would like to see estimates
that cluster tightly around the true value $\alpha = 6$
(indicated by a blue dashed line). 
The scatter in the estimates for the two methods based on harmonic analysis (top two rows)
is much larger than that for the three remaining methods.
With only a few exceptions,
this pattern persists for all 47 buoy/unit source pairings
and for all five amounts of data
and tells us that these two methods
are not competitive with the other methods.

\begin{figure}[t!]
  \noindent\includegraphics{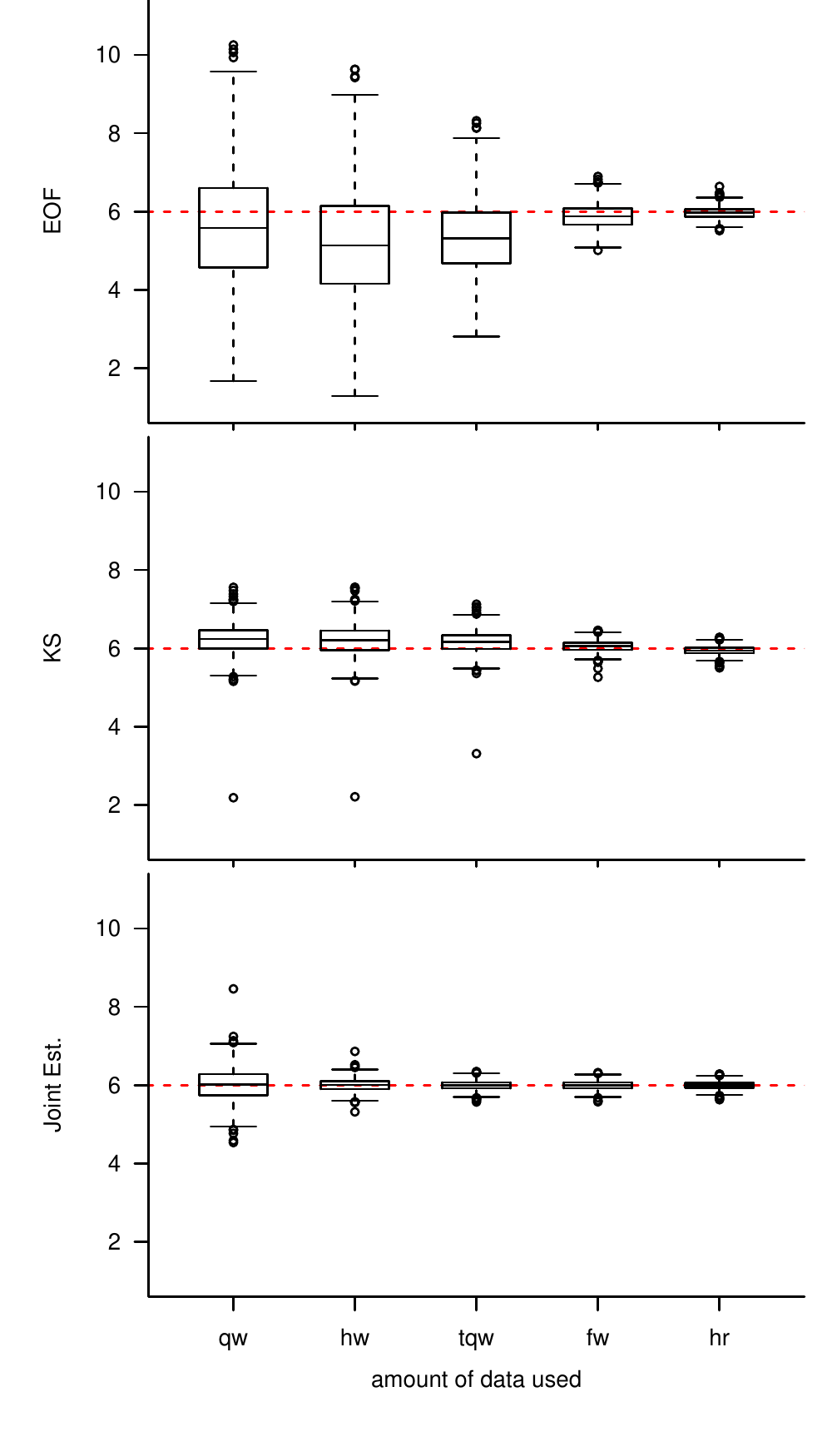}\\
  \caption{
Boxplots for 1000 estimates $\hat \alpha$
based on simulated tsunami events constructed from buoy 52402 and unit source ki060b.
The estimates are based on, respectively, 
the EOF, KS and joint estimation methods (top to bottom)
and data up to (from left to right)
a quarter of the first full wave (`qw'),
half of the full wave (`hw'), three-quarters (`tqw'),
the full wave (`fw') and one hour past the end of the first full wave (`hr').
  }
  \label{fig:boxplots}
\end{figure}

The distribution of the estimates in each row of Fig.~\ref{fig:scatterFive}
is summarized on the right-hand side by a boxplot
(Chambers {\it et al.}, 1983).  
The central box in each boxplot has three horizontal lines,
which indicate, from bottom to top,
the lower quartile, the median and the upper quartile of the data.
The short horizontal line below the central box is the lower hinge,
which indicates the estimate $\hat \alpha$
that is closest to -- but not less than --
the value of the lower quartile minus 1.5 times the interquartile distance
(the upper quartile minus the lower quartile).
The upper hinge has a similar interpretation,
with `not less than' replaced by `not greater than'
and with `lower quartile minus' replaced by `upper quartile plus'.
Any estimates $\hat \alpha$ that happen to be either smaller than the lower hinge
or greater than the upper hinge are indicated by circles.
Because the variability in the estimates for the first two methods
is so much greater than those for the three remaining methods,
only the boxplots for the former are clearly visible in Fig.~\ref{fig:scatterFive}.
Figure~\ref{fig:boxplots} shows boxplots
for the just latter three methods on a common scale,
but now for all five amounts of data rather than just
data up to 3/4 FFW.
These boxplots show that the EOF estimates tend to be biased low
when using data less than or equal to 3/4 FFW,
whereas the KS estimates are biased high, but to a lesser degree
(the boxplots show medians, but there would be no noticeable differences
had we shown sample means rather medians in Fig.~\ref{fig:boxplots}).
By contrast there is little evidence of bias in the joint-estimation estimates.
For estimates based on data up to the FFW or smaller amounts,
the spreads of the distributions are generally greatest for the EOF estimates
and smallest for the joint-estimation estimates.

\begin{figure}[t]
  \noindent\includegraphics{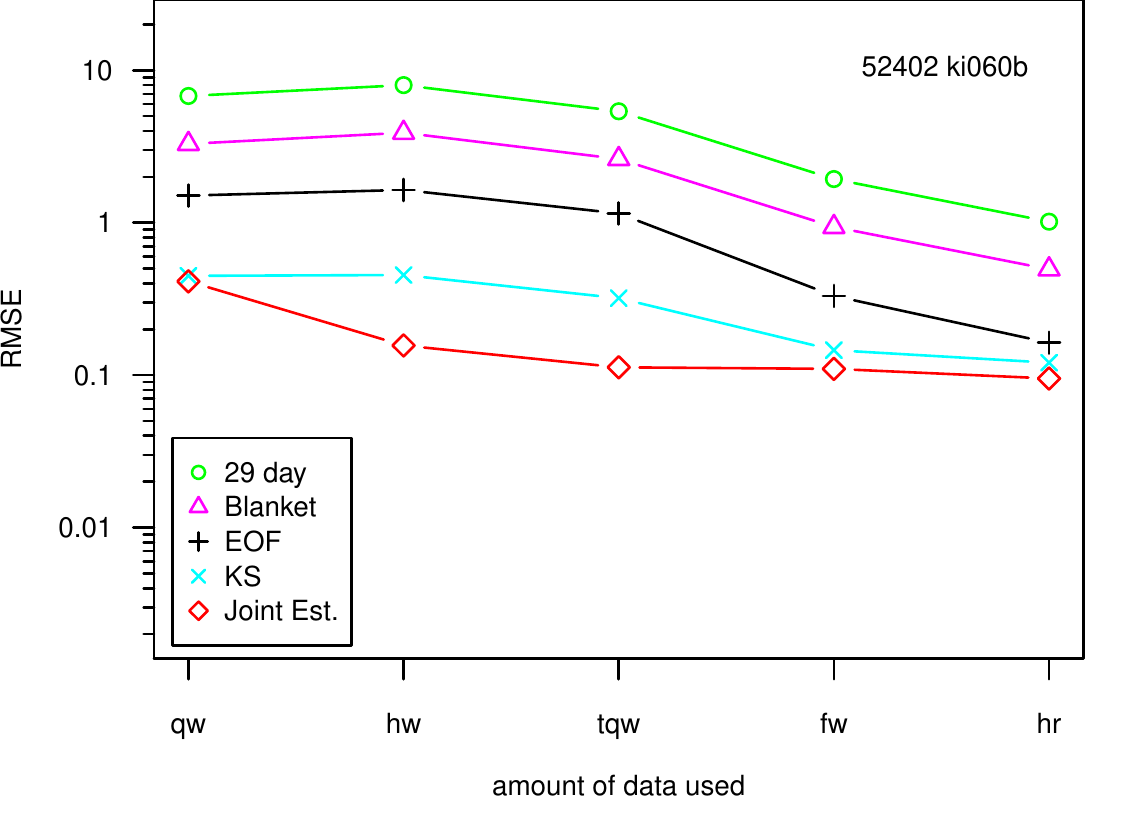}\\
  \caption{
Root-mean-square errors for 1000 estimates $\hat \alpha$
based on simulated tsunami events constructed from buoy 52402 and unit source ki060b.
For each of five methods,
RMSEs are shown for estimates based on data up to (from left to right)
a quarter of the first full wave (`qw'),
half of the full wave (`hw'),
three-quarters (`tqw'),
the full wave (`fw')
and one hour past the end of the first full wave (`hr').
  }
  \label{fig:oneRMSE}
\end{figure}

Figure~\ref{fig:oneRMSE} summarizes the spreads in the distributions
involved in Figs.~\ref{fig:scatterFive} and~\ref{fig:boxplots}
via root-mean-square errors (RMSEs), i.e.,
$\surd(\sum_i(\hat \alpha_i-6)^2/1000$).
By this measure
the two methods based on harmonic analysis are about an order of magnitude worse
than the best method (joint estimation).
With almost no exceptions,
this poor RMSE performance persists 
through all 47 buoy/unit source pairings and all five amounts of data under study.
The joint estimation method outperforms the EOF and KS methods,
but the former becomes competitive when the largest amount of data is used,
and the latter, for the smallest and two largest amounts.
Increasing the amount of data from 1/4 FFW to FFW
results in approximately half an order of magnitude drop in RMSE
for all five methods
(Table~\ref{tab:unitSources} indicates that the time it takes
to collect the extra data from 1/4 FFW to FFW is 14~min
for this particular buoy/unit source pairing).

\begin{figure}[t]
  \noindent\includegraphics{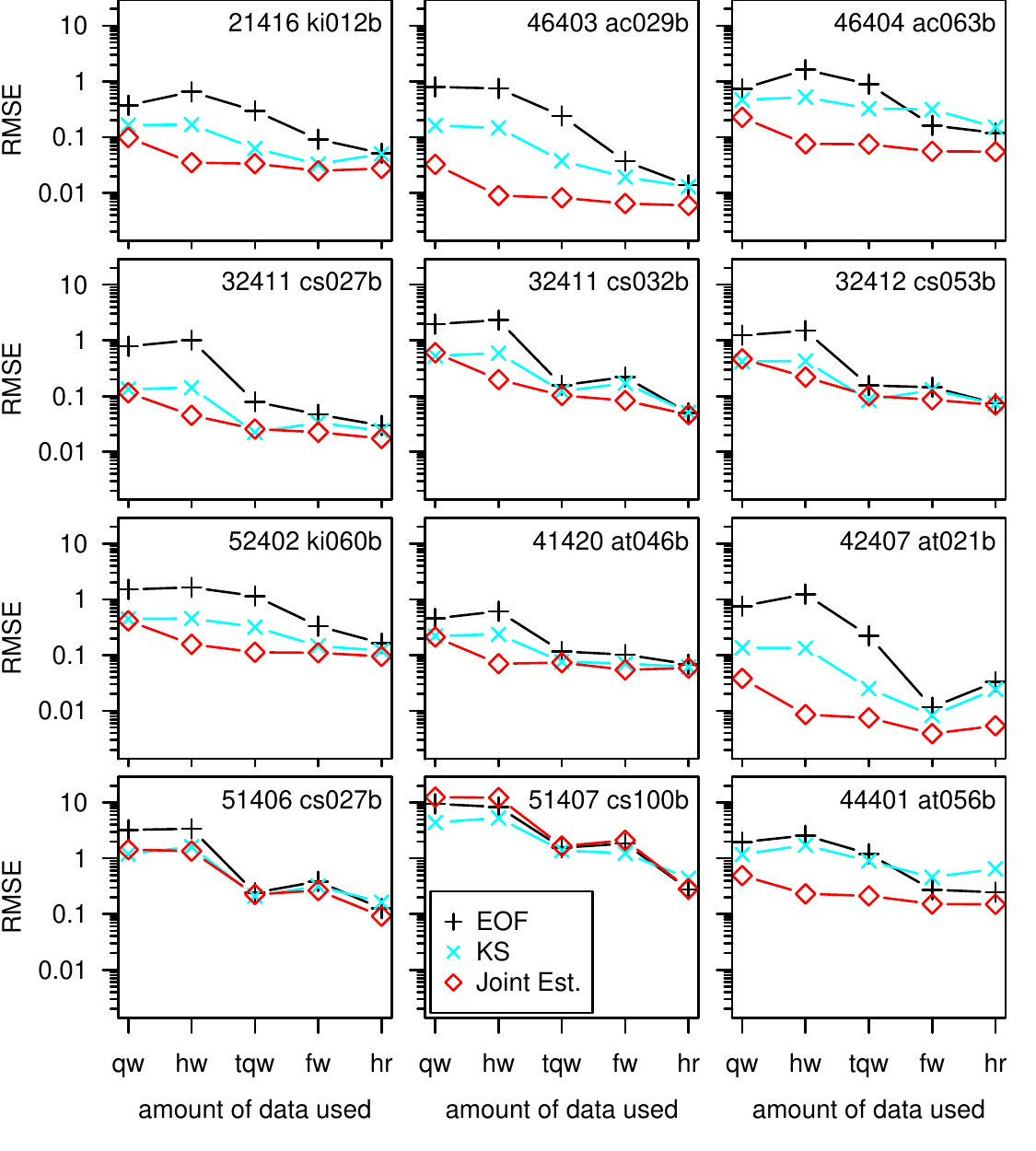}\\
  \caption{As in Fig.~\ref{fig:oneRMSE},
but now for twelve buoy/unit source pairings shown in Figs.~\ref{fig:unitSources}
and~\ref{fig:tsunamiEvents}
and with two non-competitive methods based on harmonic analysis eliminated.
  }
  \label{fig:stackedRMSEs}
\end{figure}

Figure~\ref{fig:stackedRMSEs} is similar to Fig.~\ref{fig:oneRMSE},
but now shows RMSE plots for the twelve representative buoy/unit source pairings
shown in Figs.~\ref{fig:unitSources} and~\ref{fig:tsunamiEvents}.
To simplify this figure,
we do not show results for the two non-competitive methods based upon harmonic analysis.
The joint estimation method tends to outperform the EOF and KS methods
when more than 1/4 FFW of data is involved, but not uniformly so
(the pairing of buoy 51407 with unit source cs100b in the middle plot
on the bottom row offers a counterexample).
The KS method generally outperforms the EOF method.
There are a number of instances in which RMSE increases
for the EOF and KS methods when the amount of data
increases from 1/4 FFW to 1/2 FFW.
This pattern is counterintuitive
since more data should imply a more stable estimate of $\alpha$;
however, this behavior is confirmed by an analytic theory
in which the tsunami's partial waves are regarded as a filter on the tides.

\begin{figure}[t]
  \noindent\includegraphics{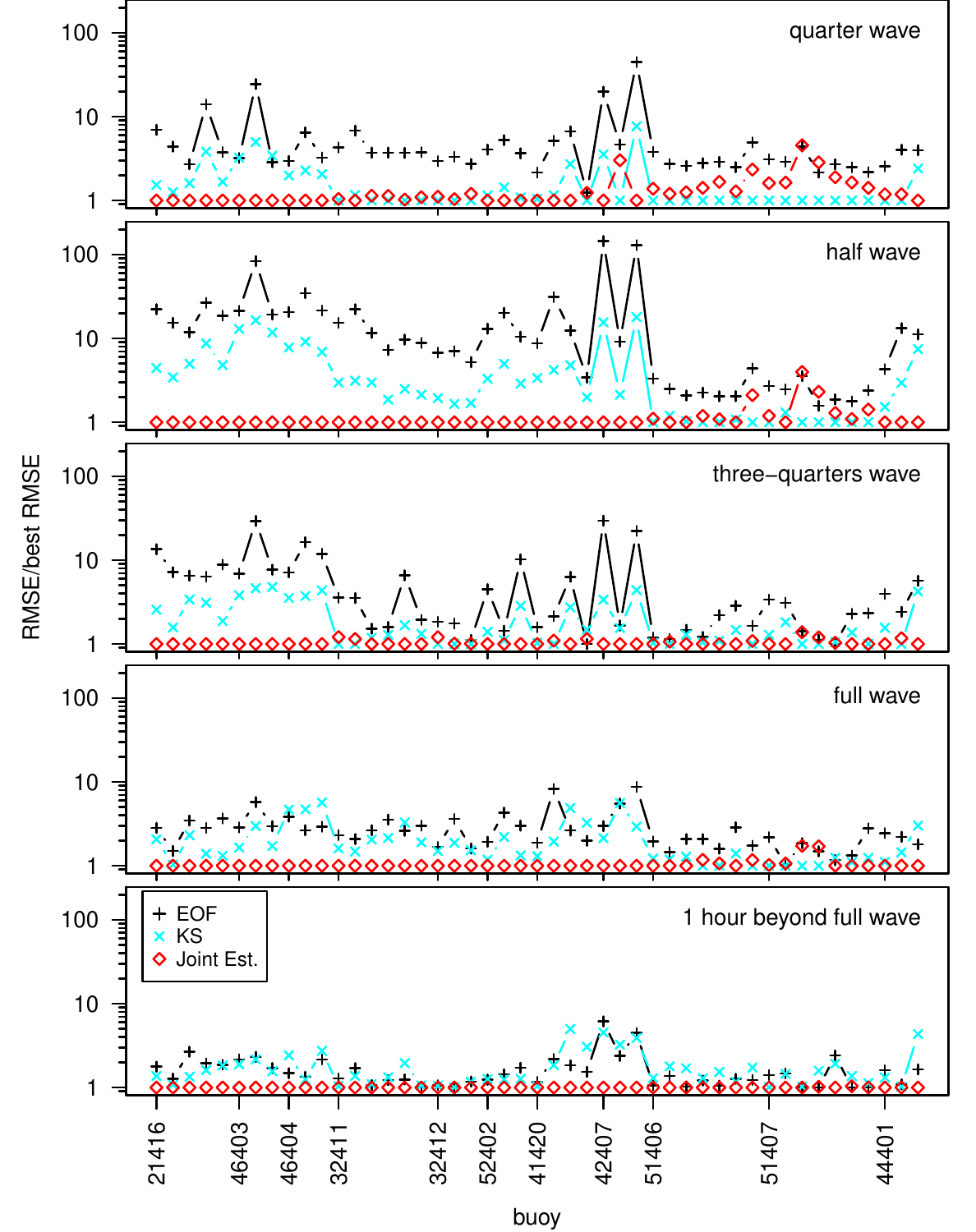}\\
  \caption{Ratio of RMSEs to best RMSE for five different amounts of data under study
(top to bottom plots), for 47 buoy/unit source pairings and
for three methods (EOF, KS and joint estimation, but with the best method chosen
from amongst these three and the two methods based on harmonic analysis).
The ordering of pairings from left to right in the plots is the same
as the ordering from top to bottom in Table~\ref{tab:unitSources}.
  }
  \label{fig:bestRMSEqw}
\end{figure}

Figure~\ref{fig:bestRMSEqw} has five plots, one for each amount of data under study,
with points indicating ratios of RMSEs involving all 47 twelve buoy/unit source pairings.
The ratios are formed by taking the RMSE for either
the EOF, KS or joint estimation method and dividing it by
the best RMSE amongst all five detiding methods. 
A ratio of one for a particular method indicates that it is the best method.
The top plot is for 1/4 FFW
and shows the KS and joint estimation methods
about evenly divided for best-method honors.
The buoys are ordered such that those that are separated most in distance
from their unit sources (51406, 51407 and 44401) are on the right-hand side
of the plots.
The KS method generally outperforms the joint estimation method for these buoys,
which tend to have lower signal-to-noise ratios
than the buoy/unit source pairings on the left-hand side of the plot.
As the amount of data increases beyond 1/4 FFW (four bottom plots),
there are increasingly fewer pairings
where the joint estimation method fails to be the method of choice.
We also note that the disparity amongst the three methods tends
to decrease as the amount of data increases.

Finally, we note that, as alternatives to RMSE as a summary measure, 
we also considered the mean absolute error $\sum_i|\hat \alpha_i-6|/1000$,
the maximum absolute error $\max\{|\hat \alpha_i-6|\}$
and the maximum absolute negative error,
i.e., $\max\{|\hat \alpha_i-6|\}$
with $i$ ranging over values such that $\hat \alpha_i < 6$. 
These last two measures are based on worst-case scenarios,
which are of considerable operational interest;
in particular, the last measure focuses on worst-case underestimation of $\alpha$,
which can lead to forecasting coastal inundations that are too small.
All three additional measures lead to the same conclusions
we drew using the RMSE:
the joint estimation method is generally to be preferred over
the EOF and KS methods,
and the two methods involving only harmonic analyses are not competitive.

\section{Example: March 2011 Japan tsunami}\label{sec:Tohoku}
\begin{figure}[t!]
  \noindent\includegraphics{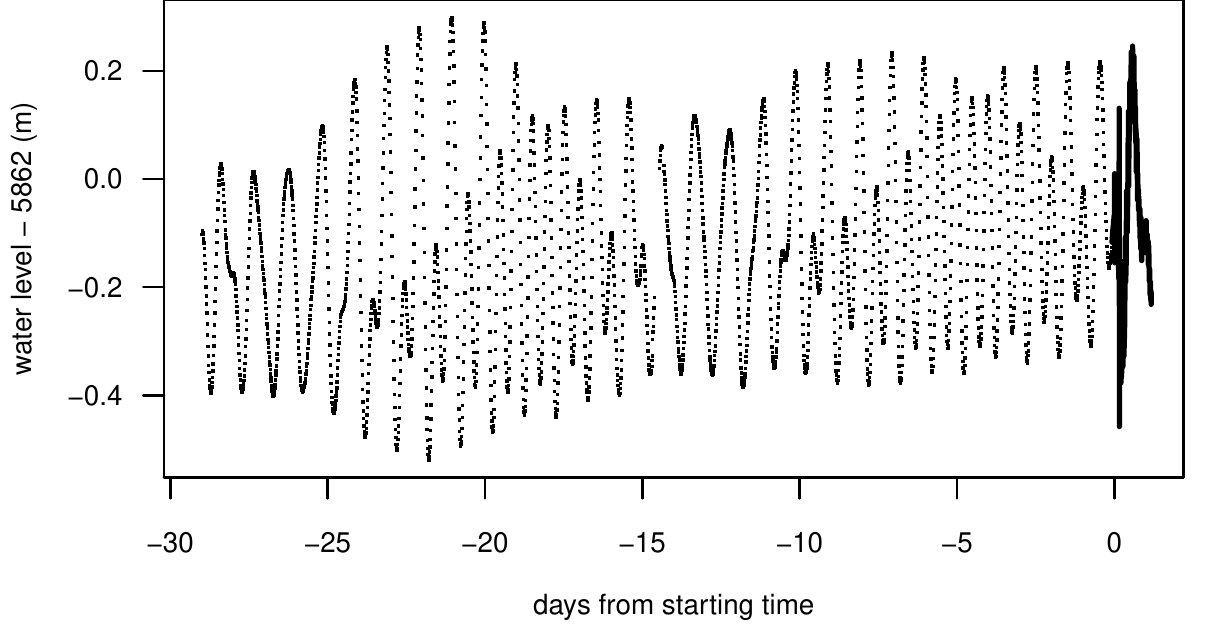}\\
  \caption{
Data recorded by \DART\ buoy 52402 during March 2011 Japan tsunami
with starting time $t_0$ set to 05:46:23~UT on 3/11/2011.
The points show the 15-min stream,
which goes from 29 days up to 2.0 hours prior to $t_0$.
The dark curve on the right-hand side is the 1-min stream,
which starts 1.8 hours prior to $t_0$ and ends one day after $t_0$.
  }
  \label{fig:TohokuData}
\end{figure}

Here we apply the five detiding methods
to data collected during the devastating 2011 Japan tsunami,
which was generated by the great Mw 9.0 earthquake that occurred
on March 3rd at 05:46:23~UT ($t_0$)
(Tang {\it et al.}, 2012; Wei {\it et al.}, 2012; Wei {\it et al.}, 2014).
Several \DART\ buoys recorded this event,
including buoy 52402.
Figure~\ref{fig:TohokuData} shows the data
from 52402 we make use of here.
The data consist of a 15-min stream
starting 29~days prior to $t_0$
and ending 2.0~hrs before $t_0$,
and a 1-min stream starting 1.8~hrs prior to $t_0$
and ending a day after $t_0$
(there is a small gap in the 15-min stream
around 14 days prior to $t_0$).

\begin{figure}[t]
  \noindent\hskip-6pt\includegraphics{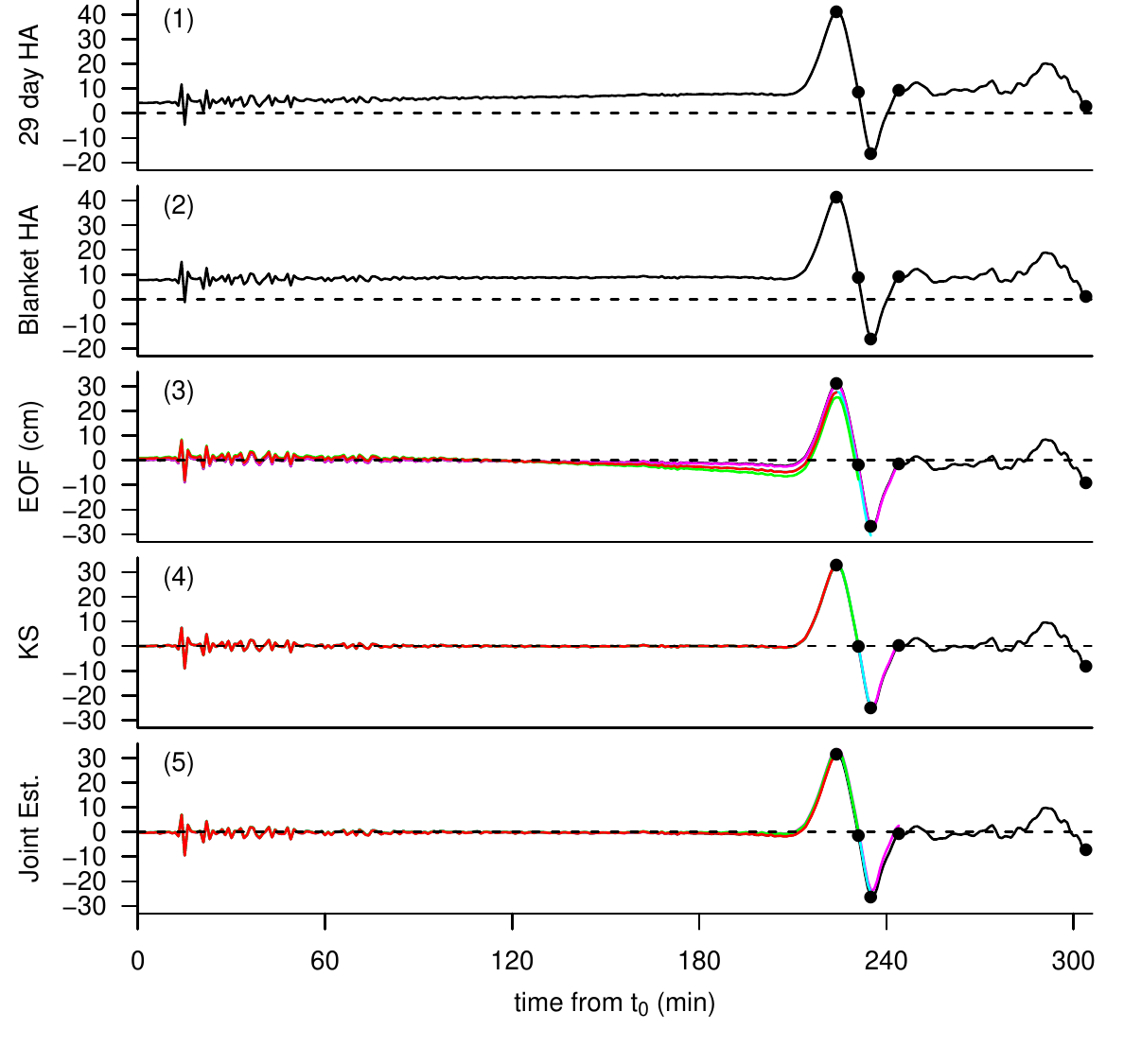}\\
\caption{
Detided bottom pressure measurements for March 2011 Japan tsunami data
shown in Fig.~\ref{fig:TohokuData}.
Each plot corresponds to a different detiding method.
From top to bottom,
these are
(1)~a 6 constituent harmonic analysis (HA)
based on 29 days of data before $t_0$ (the start of the event),
(2)~a 68 constituent HA based on 15-sec data
recorded from 12/13/2006 to 3/21/2008;
(3)~an empirical orthogonal function approach,
(4)~a Kalman smoother approach
and
(5)~a 2 constituent local HA with joint estimation of source coefficients.
See the caption to Fig.~\ref{fig:detidingFive}
for an explanation of the black circles and the color-coded curves
in the bottom three plots. 
}
  \label{fig:detidingTohoku}
\end{figure}

The first method uses a 6 constituent harmonic analysis
based on the 29 days of data prior to the event time $t_0$
(these are mostly from the 15-min stream,
but there are 1.8 hours from the 1-min stream).
The top plot in Fig.~\ref{fig:detidingTohoku}
shows the corresponding detided series,
which is the difference between data from the 1-min stream recorded
after $t_0$ and predicted values based on the fitted harmonic model.
This series shows rapid fluctuations starting about 10~min after $t_0$
and dissipating after about 90~min.
These are evidently due to seismic noise from the earthquake.
Ignoring this noise,
the detided series starts at a positive intercept and
then rises almost linearly until about 210 minutes,
at which point the tsunami signal becomes evident.
The five black circles are subjectively chosen markers
of a quarter of the first full wave (FFW), a half, three quarters,
the end of the FFW and one hour past the end of the FFW
(these occur at, respectively, 225, 232, 236, 245 and 305~min after $t_0$). 
The true tidal fluctuations are of course unknown,
but a reasonable conjecture is that
the linear increase in evidence here is actually
a tidal component that the first method failed to properly extract.

The second method uses a 68 constituent harmonic analysis
based on 465 days of data collected by buoy 52402
between 12/13/2006 and 3/21/2008,
i.e., well before the 2011 Japan tsunami.
The detided series,
shown below the top plot in Fig.~\ref{fig:detidingTohoku},
is again the difference between data from the 1-min stream
and predicted values based on the fitted harmonic model,
but with the mean level of the predictions
adjusted using the 29 days of data prior to $t_0$.
In comparison to the first method,
there is now only a slight linear increase in the first three hours,
but the positive intercept is larger.

The third method is EOF-based and yields detided series
that depend on the amount of the 1-min stream
after $t_0$ we wish to detide.
The middle plot in Fig.~\ref{fig:detidingTohoku}
shows five detided series color coded
to indicate the amount of data used
(red, green, cyan, magenta and black
for data ending at, respectively from left to right, 
the locations of the five solid circles,
and starting 1470 minutes earlier).
Differences between the displayed detided series
at the same location in time
are typically small, but do get as large as 5.7~cm.
Ignoring seismic noise,
all five detided series are relatively flat
out to 150~min,
after which they exhibit a noticeable dip
prior to the arrival of the tsunami signal.
While we might be tempted to regard this dip as a remanent tidal component,
caution is in order since true tidal fluctuations are unknown.

The fourth method -- Kalman smoothing --
is similar to the EOF method
in that the detided series depends on the amount of the 1-min stream to be detided,
but is dissimilar in that detiding involves data prior to $t_0$
only indirectly because the starting point for the KS method
is the detided series provided by the first method.
The next-to-bottom plot in Fig.~\ref{fig:detidingTohoku}
shows five detided series color coded in the same manner as before.
The differences between the detided series
at the same point in time are again typically small,
with the largest difference now being 0.7~cm.
There is only a slight hint here
of the dip readily apparent in the EOF detided series.

Similar to the EOF and KS methods,
the final method (a 2 constituent local harmonic analysis
with joint source coefficient estimation)
yields detided series that depend upon the amount of the 1-min stream to be detided,
but, in contrast, it makes no direct or indirect use of any data prior to $t_0$.
This method, however, does depend upon a suitable model for the tsunami signal.
Percival {\it et al.}~(2014) discuss selection of unit sources for the Japan tsunami
using an objective automatic method.
The selection is based on data from three \DART\ buoys (21401, 21413 and 21418)
located much closer to the epicenter of the earthquake than buoy 52402 is.
These buoys registered the tsunami signal within 5~min after $t_0$,
well before it arrived at buoy 52402 more than 3~h later.
The automatic method selected seven unit sources ${\bf g}_1$, \dots, ${\bf g}_7$ to model the tsunami signal.
Accordingly, we need to adjust the joint estimation method
to make use of seven unit sources rather than just one.
We do so by suitably redefining the design matrix $X$
and the vector of coefficients $\bm \alpha$ in Equation~\ref{eq:jointRewritten}.
Thus $X$ is now of dimension $N\times 12$,
with its first five columns being as before
and with its next seven columns now being ${\bf g}_1$, \dots, ${\bf g}_7$;
correspondingly,
$\bm \beta$ is augmented to dimension 12,
with its last seven elements being the source coefficients for the unit sources.
After these adjustments to $X$ and $\bm \beta$,
we produce the detided series using the same equations as before 
(\ref{eq:normalEquations} and \ref{eq:jointDetided}).
The bottom plot in Fig.~\ref{fig:detidingTohoku}
shows detided series for the joint estimation method
(five in all corresponding to different amounts of data).
These series are visually quite similar to the KS detided series,
but the largest difference between the joint detided series
is larger (3.6~cm) than for the KS series (0.7~cm).

It is important to note that,
in contrast to the other four methods,
the joint estimation method is dependent upon a suitable model for the tsunami signal.
To demonstrate this fact,
the two plots in Fig.~\ref{fig:detidingTohokuOneUS}
show detided series using the joint estimation method, but based upon different models.
The top plot is the same as the bottom plot in Fig.~\ref{fig:detidingTohoku},
for which the detided series utilize a model involving seven unit sources.
In the bottom plot the detided series use a model based on only one of these seven sources (ki026b).
The magnitude of the earthquake that generated the 2011 Japan tsunami
was so large that it is physically unrealistic for the signal
to be well-modeled by a single unit source
(Papazachos {\it et al.}, 2004).
The detided series from this presumably inadequate model
have low-frequency fluctuations in the first three hours
that are not evident in the top plot
(or in the KS-based detided series,
which are shown in the next-to-bottom plot of Fig.~\ref{fig:detidingTohoku}).
These fluctuations are best attributed to a failure
on the part of the joint estimation method
due to an inappropriate model. 

\section{Conclusions and discussion}\label{sec:conclusions}
We have undertaken a comprehensive comparison of five methods
for estimating the source coefficient $\alpha$ based upon \DART\ buoy 
bottom pressure (BP) data collected during a tsunami event
(this coefficient reflects the strength of the event and
is used to provide initial conditions for predicting coastal inundation).
Any method for estimating $\alpha$ must deal with the fact
that the variability in BP data is typically dominated by tidal fluctuations,
and hence all viable methods must detide the data in some explicit or implicit manner.
The five methods under study have been entertained as part of the on-going
development of the SIFT application,
a tool developed at NOAA for use by U.S.~Tsunami Warning Centers
for real-time assessment of tsunami events.

The clear method of choice is a scheme by which $\alpha$ is estimated
jointly in a regression model that accounts for the tidal components
using sinusoidal constituents involving the tidal frequency M2 and half that frequency.
This method is particularly convenient from an operational point of view
in that it does not make direct use of data occurring prior to a tsunami event,
as is true in varying degrees for the four other methods under study.
Amongst the four remaining methods,
the Kalman smoothing (KS) method performed best overall in our study.
In some cases,
the EOF method is competitive with the joint estimation and KS methods,
but the two methods based on only harmonic analyses proved to be markedly inferior
to the other three methods. 
Note that we evaluated the performance of the EOF method
using a specific set of eight basis vectors
adapted for use within the SIFT system.
We limited this set to eight vectors
due to the fact that only this number of vectors
defines a location-independent tidal sub-space,
should the set be derived using data from a single buoy,
as was the case here (Tolkova, 2010).  
Expanding the set of vectors by deriving them
from multiple buoys might allow for more accurate detiding.

\begin{figure}[t]
  \noindent\hskip-6pt\includegraphics{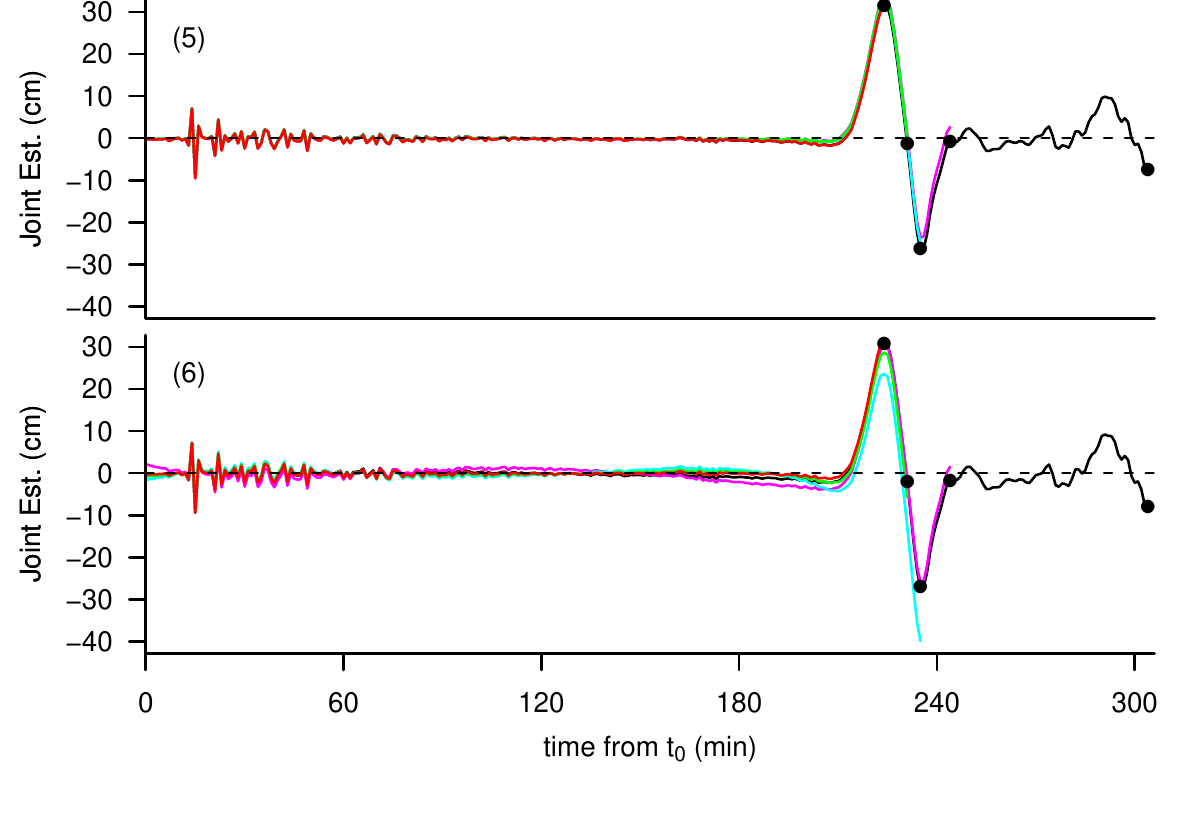}\\
\caption{
Detided bottom pressure measurements for March 2011 Japan tsunami data
shown in Fig.~\ref{fig:TohokuData}
using the method based on a two constituent local harmonic analysis
with joint estimation of source coefficients.
The top plot here is identical to the bottom plot of Fig.~\ref{fig:detidingTohoku},
for which the detided series are based on a model for the tsunami signal
that is the linear combination of six unit sources selected by an objective automatic procedure
(Percival {\it et al.}, 2014).
In the bottom plot, the detided series are based on a single unit source 
that is presumably an inadequate description of the underlying tsunami signal.
See the caption to Fig.~\ref{fig:detidingFive}
for an explanation of the black circles and the color-coded curves in these plots. 
}
  \label{fig:detidingTohokuOneUS}
\end{figure}

Despite the fact that our study points to the joint estimation method
as the method of choice, some caution is in order.
A presumption behind our study is that
the model for the tsunami signal is known perfectly,
which is obviously unrealistic in practice.
The issue of model mismatch must therefore temper our conclusions,
a point that is reinforced by the discussion surrounding Fig.~\ref{fig:detidingTohokuOneUS}. 
For graphical presentation of detided series within SIFT,
the fact that the KS method does not depend on an assumed model for the tsunami signal
-- but often compares favorably with the joint estimation method --
suggests its use.
As currently implemented,
the KS method uses the output from the 29-day harmonic analysis method
as its starting point.
In cases where the 15-min stream is not available over much of the preceding 29 days,
the blanket harmonic method could provide a model-free detiding
for display purposes,
as could the EOF method if the 15-min stream were available going back
about a lunar day from the start of the tsunami signal. 
There is thus potential use within SIFT for all five methods we have studied.
As of this writing, the current version of SIFT
supports the 29-day harmonic analysis, EOF and KS methods,
and there are plans to implement the joint estimation method
within an overall scheme for automatically selecting unit sources
to serve as models for the tsunami signal.

\begin{acknowledgements} 
This work was funded by the Joint Institute for the Study of the Atmosphere and Ocean (JISAO)
under NOAA Cooperative Agreement No.~NA17RJ1232 and is JISAO Contribution No.~2185. 
This work is also Contribution No.~4089 from NOAA/Pacific Marine Environmental Laboratory.
The authors thank George Mungov of NOAA's National Geophysical Data Center
for supplying \DART\ buoy data and predictions based upon harmonic analyses with 68 sinusoidal constituents.
\end{acknowledgements}


\newpage
\begin{table}
\begin{center}
\begin{tabular}{c|c|c|c|c|c}
&&tidal&&&\\
buoy&location&range&type of tide&start/stop&days\\
\hline\hline
21416&North Pacific     &1.28 m&mixed-diurnal&07/25/2007&708\\&&&&07/01/2009\\
46403&North Pacific     &2.78 m&mixed-semidiurnal&08/21/2007&423\\&&&&10/16/2008\\
46404&North Pacific     &3.26 m&mixed-semidiurnal&12/06/2007&541\\&&&&05/29/2009\\
32411&East Pacific      &1.70 m&semidiurnal&03/27/2007&766\\&&&&04/30/2009\\
32412&East Pacific      &0.67 m&mixed-diurnal&11/01/2007&709\\&&&&10/09/2009\\
52402&West Pacific      &0.67 m&mixed-diurnal&12/13/2006&465\\&&&&03/21/2008\\
41420&West Atlantic     &0.84 m&mixed-semidiurnal&06/25/2008&321\\&&&&05/11/2009\\
42407&West Atlantic     &0.24 m&diurnal&04/10/2006&756\\&&&&05/04/2008\\
51406&mid-ocean Pacific &1.87 m&semidiurnal&09/04/2006&623\\&&&&05/18/2008\\
51407&mid-ocean Pacific &0.83 m&mixed-diurnal&08/08/2007&728\\&&&&08/04/2009\\
44401&mid-ocean Atlantic&0.95 m&mixed-semidiurnal&08/31/2007&998\\&&&&05/24/2010\\
\hline\hline
\end{tabular}
\end{center}
\caption{Eleven representative \DART\ buoys used in detiding study.
The buoy locations are shown by triangles on Fig.~\ref{fig:DetidingPaperChart}.
The tidal range is a mean daily range as estimated from the harmonic constituents
described in Sect.~\ref{sec:Harmonic}
for the second harmonic method (Mofjeld {\it et al.}, 2004).
The type of tide is also based upon these constituents (Parker, 2007).
The start/stop dates are those associated
with the bottom pressure data retrieved from the particular buoy,
and the column marked `days' gives the number of days spanned by the data. 
}
\label{tab:buoys}
\end{table}
\clearpage
\begin{table}
\begin{center}
\begin{tabular}{c|ccccccl}
buoy&unit source&$1/4$&$1/2$&$3/4$&full&period (min)&range (cm)\\
\hline\hline
21416&ac005b                  &  54 &  56 &  57 &  64 & 13 &\hbox to 27pt{\hfil4.6}\\
&ac010b                       &  69 &  72 &  74 &  77 & 11 &\hbox to 27pt{\hfil2.2}\\
&ki002b                       &  49 &  54 &  56 &  63 & 19 &\hbox to 27pt{\hfil1.6}\\
&ki007b                       &  29 &  31 &  32 &  36 &  \,\,\,9 &\hbox to 27pt{\hfil11.8}\\
&$\hbox{ki012b}^\dagger$       &  46 &  49 &  52 &  55 & 12 &\hbox to 27pt{\hfil3.1}\\
46403&ac026b                  &  23 &  26 &  30 &  35 & 16 &\hbox to 27pt{\hfil3.5}\\
&$\hbox{ac029b}^\dagger$       &  13 &  15 &  17 &  21 & 11 &\hbox to 27pt{\hfil13.1}\\
& ac032b                      &  30 &  35 &  37 &  42 & 16 &\hbox to 27pt{\hfil2.9}\\
46404&ac055b                  &  41 &  46 &  50 &  73 & 43 &\hbox to 27pt{\hfil2.4}\\
& ac059b                      &  27 &  29 &  31 &  37 & 13 &\hbox to 27pt{\hfil9.0}\\
&$\hbox{ac063b}^\dagger$       &  43 &  49 &  53 &  68 & 33 &\hbox to 27pt{\hfil2.3}\\
32411&cs019b                  &  82 &  85 &  88 &  91 & 12 &\hbox to 27pt{\hfil2.7}\\
&$\hbox{{cs027b}}^\dagger$     &  67 &  69 &  72 &  74 &  \,\,\,9 &\hbox to 27pt{\hfil5.3}\\
&$\hbox{cs032b}^\dagger$       & 103 & 106 & 111 & 114 & 15 &\hbox to 27pt{\hfil2.0}\\
& cs040b                      & 144 & 147 & 152 & 156 & 16 &\hbox to 27pt{\hfil3.0}\\
& cs045b                      & 127 & 131 & 135 & 153 & 35 &\hbox to 27pt{\hfil1.0}\\
& {cs049b}                    & 148 & 151 & 155 & 158 & 13 &\hbox to 27pt{\hfil3.4}\\
32412&$\hbox{{cs053b}}^\dagger$& 102 & 105 & 108 & 112 & 13 &\hbox to 27pt{\hfil1.4}\\
& cs061b                      &  79 &  81 &  84 &  87 & 11 &\hbox to 27pt{\hfil4.0}\\
& cs070b                      & 121 & 124 & 128 & 131 & 13 &\hbox to 27pt{\hfil1.6}\\
52402&ki050b                  &  70 &  73 &  78 &  83 & 17 &\hbox to 27pt{\hfil1.5}\\
& {ki055b}                    &  57 &  59 &  61 &  63 & \,\,\,8 &\hbox to 27pt{\hfil5.2}\\
&$\hbox{ki060b}^\dagger$       &  78 &  81 &  83 &  92 & 19 &\hbox to 27pt{\hfil0.8}\\
41420&$\hbox{at046b}^\dagger$  &  50 &  52 &  57 &  59 & 12 &\hbox to 27pt{\hfil1.6}\\
& at051b                      &  31 &  33 &  35 &  37 & \,\,\,8 &\hbox to 27pt{\hfil7.5}\\
& {at056b}                    &  53 &  59 &  65 &  88 & 47 &\hbox to 27pt{\hfil1.1}\\
& at092b                      &  42 &  61 &  68 &  74 & 43 &\hbox to 27pt{\hfil0.2}\\
42407&$\hbox{at021b}^\dagger$  &  19 &  21 &  23 &  26 & \,\,\,9 &\hbox to 27pt{\hfil10.6}\\
& at037b                      & 109 & 114 & 127 & 132 & 31 &\hbox to 27pt{\hfil0.2}\\
& at090b                      &  18 &  19 &  21 &  24 & \,\,\,8 &\hbox to 27pt{\hfil10.6}\\
51406&cs004b                  & 314 & 317 & 322 & 325 & 15 &\hbox to 27pt{\hfil0.9}\\
&$\hbox{{{cs027b}}}^\dagger$   & 412 & 416 & 421 & 425 & 17 &\hbox to 27pt{\hfil0.8}\\
& {{cs053b}}                  & 404 & 408 & 414 & 419 & 20 &\hbox to 27pt{\hfil0.5}\\
& cs080b                      & 513 & 517 & 523 & 528 & 20 &\hbox to 27pt{\hfil0.4}\\
& cs104b                      & 567 & 572 & 576 & 588 & 28 &\hbox to 27pt{\hfil0.8}\\
& nt008b                      & 518 & 523 & 526 & 530 & 16 &\hbox to 27pt{\hfil0.3}\\
& nt034b                      & 408 & 412 & 417 & 422 & 19 &\hbox to 27pt{\hfil0.2}\\
51407&ac020b                  & 285 & 289 & 292 & 294 & 12 &\hbox to 27pt{\hfil0.3}\\
& ac065b                      & 301 & 312 & 315 & 322 & 28 &\hbox to 27pt{\hfil0.2}\\
& {{cs049b}}                  & 718 & 723 & 730 & 735 & 23 &\hbox to 27pt{\hfil0.2}\\
&$\hbox{cs100b}^\dagger$       & 919 & 929 & 938 & 942 & 31 &\hbox to 27pt{\hfil0.6}\\
& ki005b                      & 385 & 396 & 405 & 412 & 36 &\hbox to 27pt{\hfil0.2}\\
& {{ki055b}}                  & 454 & 461 & 474 & 478 & 32 &\hbox to 27pt{\hfil0.2}\\
& nt037b                      & 326 & 330 & 332 & 335 & 12 &\hbox to 27pt{\hfil0.9}\\
44401&at035b                  & 130 & 134 & 138 & 142 & 16 &\hbox to 27pt{\hfil0.3}\\
& at045b                      &  59 &  61 &  63 &  66 & \,\,\,9 &\hbox to 27pt{\hfil2.8}\\
&$\hbox{{{at056b}}}^\dagger$   &  81 &  88 &  94 & 115 & 45 &\hbox to 27pt{\hfil0.8}\\
\hline\hline
\end{tabular}
\end{center}
\caption{Unit sources used with eleven \DART\ buoys.
There are 47 pairings of buoys with unit sources;
however, only 42 distinct unit sources are involved
because five of them (cs027b, cs049b, cs053b, ki055b and at056b) 
are associated with two buoys.
The locations of the 42 unit sources
are shown by solid circles on Fig.~\ref{fig:DetidingPaperChart}.
Each buoy/unit source pairing leads to an artificial tsunami signal $\bf g$.
The columns labeled as `$1/4$', `$1/2$',  `$3/4$' and `full'
give the approximate number of minutes from $t_0$ that pass
until a quarter of the first full wave, a half, three-quarters and the first full wave
of the signal appear (these points are hand-picked).
The `period' column gives a rough measure of the duration of the first full wave
(it is equal to $4/3$ times the difference between the full-wave and quarter-wave points).
The `range' column gives the difference between the largest and smallest values
in the signal $\bf g$.
The twelve buoy/unit source pairings that lead to the signals
shown in Figs.~\ref{fig:unitSources} and~\ref{fig:tsunamiEvents}
are indicated by daggers. 
}
\label{tab:unitSources}
\end{table}
\clearpage
\begin{table}
\begin{center}
\begin{tabular}{c|ccccc}
method&$1/4$&$1/2$&$3/4$&full&full$+$1 hour\\
\hline\hline
29 day HA &5.04&4.95&5.28&5.71&5.69\\
Blanket HA&9.25&9.77&8.55&6.90&6.28\\
EOF       &7.72&7.16&6.70&6.24&5.77\\
KS        &6.29&6.22&6.18&6.06&5.88\\
Joint Est.&6.03&6.02&5.98&5.99&6.02\\
\hline\hline
\end{tabular}
\end{center}
\caption{Estimates $\hat \alpha$ of source coefficient
based on detided bottom pressure measurements
shown in Fig.~\ref{fig:detidingFive}.
There are five different estimates for each method.
These estimates correspond to use of different amounts of data,
which are indicated by black circles in Fig.~\ref{fig:detidingFive}.
and described in the caption to that figure. 
Agreement between $\hat\alpha$ and $\alpha=6$ tends to get better
as the amount of data increases.
The fifth method
(`Joint Est.', a 2 constituent local harmonic analysis
with joint estimation of the source coefficient $\alpha$)
does best overall in this example. 
}
\label{tab:alphaEstimates}
\end{table}
\clearpage
\clearpage
\clearpage
\clearpage 
\clearpage 
\clearpage
\clearpage
\clearpage
\clearpage
\clearpage
\clearpage
\clearpage
\clearpage
\clearpage
\clearpage
\end{document}